\documentclass[10 pt,letterpaper,showkeys]{article}
\usepackage{graphicx}
\usepackage{amssymb,amsmath,mathrsfs}
\usepackage{amsthm,enumerate,verbatim}
\newtheorem{theorem}{Theorem}[section]
\newtheorem{lemma}{Lemma}[section]
\newtheorem{corollary}{Corollary}[section]
\newtheorem{proposition}{Proposition}[section]
\newtheorem{remark}{Remark}[section]
\newtheorem{assumption}{Assumption}[section]

\newcommand{\beq}{\begin{equation}}
\newcommand{\eeq}{\end{equation}}
\newcommand{\beqa}{\begin{eqnarray}}
\newcommand{\eeqa}{\end{eqnarray}}
\newcommand{\beqas}{\begin{eqnarray*}}
\newcommand{\eeqas}{\end{eqnarray*}}
\newcommand{\ba}{\begin{array}}
\newcommand{\ea}{\end{array}}
\newcommand{\bi}{\begin{itemize}}
\newcommand{\ei}{\end{itemize}}
\newcommand{\gap}{\hspace*{2em}}

\newcommand{\nn}{\nonumber}

\def\vgap{\vspace*{.1in}}
\def\QED{\ifhmode\unskip\nobreak\fi\ifmmode\ifinner\else\hskip5pt\fi\fi
  \hbox{\hskip5pt\vrule width5pt height5pt depth1.5pt\hskip1pt}}

\newcommand{\cS}{{\mathcal{S}}}

\newcommand{\cI}{{\mathcal{I}}}
\newcommand{\cJ}{{\mathcal{J}}}
\newcommand{\cX}{{\mathcal{X}}}
\newcommand{\cM}{{\mathcal{M}}}

\def\balpha{{\bar \alpha}}
\def\barf{{\bar f}}
\def\D{{\mathscr D}}

\def\tr{{\rm tr}}
\def\tw{{\tilde w}}
\def\vec{\mathop{{\bf vec}}}
\def\eps{{\epsilon}}
\def\Null{{\rm Null}}
\def\range{{\rm Range}}
\def\svec{{\bf svec}}
\def\svecr{\svec_0}
\def\smat{\mathop{{\bf smat}}}
\def\sotimes{{\otimes_s}}

\title{Computing Optimal Experimental Designs via Interior Point Method
\thanks{This work was supported in part by an NSERC Discovery Grant.}
}

\author{
    Zhaosong Lu%
    \thanks{
    Department of Mathematics, Simon Fraser University, Burnaby, BC,
    V5A 1S6, Canada. (email: {\tt zhaosong@sfu.ca}).}
    \and
    Ting Kei Pong
    \thanks{Department of Combinatorics and Optimization, University of Waterloo,
    Waterloo, ON, N2L 3G1, Canada. (email: {\tt ptingkei@math.uwaterloo.ca}).}
}

\date{October 12, 2012}

\begin{document}

\maketitle

\begin{abstract}

In this paper, we study optimal experimental design problems with a broad class of
smooth convex optimality criteria, including the classical A-, D- and $p$th mean
criterion. In particular, we propose an interior point (IP) method for them and
establish its global convergence. 
Furthermore, by exploiting the structure of the Hessian matrix of the aforementioned optimality criteria,  
we derive an explicit formula for computing its rank.
Using this result, we then show that the Newton direction arising in the IP method
can be computed efficiently via Sherman-Morrison-Woodbury formula when the
size of the moment matrix is small relative to the sample size. Finally, we compare
our IP method with the widely used multiplicative algorithm introduced by Silvey et
al. \cite{STT78}. The computational results show that the IP method generally
outperforms the multiplicative algorithm both in speed and solution
quality. 

\vskip14pt


\vskip14pt


\end{abstract}

\section{Introduction}
\label{intro}

In this paper, we consider the optimal experimental design
problems on a given finite design space $\cX = \{x_1,\ldots,x_n\}\subseteq \Re^m$.
In this setting, we consider a coefficient matrix $K\in \Re^{m\times k}$ of full column rank and the moment matrix defined as
\[
\cM(w)=\sum_{i=1}^nw_iA_i
\]
for $w\in \Omega:=\{w:\;w_i\ge 0,\sum_{i=1}^nw_i=1\}$, where $A_i$ is the expected Fisher information matrix related to
$x_i$, $i=1,...,n$. As in \cite{Yu10a}, throughout this paper we assume that $A_i$'s are $m \times m$ real symmetric
positive semidefinite matrices and that there exists an $w\in \Omega$ such that $\cM(w)$ is positive definite.
This in particular implies that
$\cM(w)$ is positive definite for all positive $w\in \Omega$.
The optimal experimental design problem can then be
formulated as the following minimization problem (see \cite[Section~7.10]{Puk93}):
\begin{equation}\label{design}
  \begin{array}{rl}
    f^*:=\inf\limits_w & \Phi(\cM(w)) := \Psi({\cal C}_K(\cM(w)))\\
    {\rm s.t.} & w\in \Omega, \ {\rm Range}(K)\subseteq {\rm Range}(\cM(w)),
  \end{array}
\end{equation}
where $\Psi$ is a function defined on the set of positive
definite matrices and ${\cal C}_K(\cM(w))$ is the information matrix
defined by ${\cal C}_K(\cM(w)):=(K^T(\cM(w))^\dagger K)^{-1}$. Here $A^\dagger$ denotes the Moore-Penrose
pseudoinverse of a matrix $A$. The well-definedness of ${\cal C}_K(\cM(w))$ is guaranteed by the range inclusion
condition in the constraint of \eqref{design}
and the fact that $K$ has full column rank \cite[Chapter~3]{Puk93}.
The function $\Phi$ in the objective is commonly referred to as an ``optimality criterion".
Some classical optimality criteria include (see \cite[Chapter~6]{Puk93}):
\begin{enumerate}[(i)]
  \item A-criterion $\Phi(X):=\tr(K^TX^\dagger K)$;
  \item c-criterion $\Phi(X):=c^TX^\dagger c$;
  \item D-criterion $\Phi(X):=\log\det(K^TX^\dagger K)$;
  \item $p$th mean criterion $\Phi(X):=
    \tr((K^TX^\dagger K)^{-p})$.
\end{enumerate}
for some $p<0$, $c\in \Re^m$ and $K \in \Re^{m\times k}$ of full column rank.

It is easy to observe that c-criterion is just a special case of A-criterion
with $K=c$ and A-criterion is a special case of $p$th mean
criterion with $p=-1$. We shall also mention that $p$th mean criterion
can be defined more generally to include D-criterion as a
special case (see \cite[Chapter~6]{Puk93} for details). Furthermore,
it can be shown that the constraint set of \eqref{design} is convex \cite[Section~3.3]{Puk93},
and the criteria (i)-(iv) are convex functions in the constraint set
(by using \cite[Theorem~5.14]{Puk93} and \cite[Theorem~6.13]{Puk93}, or
\cite[Proposition~IV.14]{Paz86} and \cite[Proposition~IV.15]{Paz86}).
Hence, problem
\eqref{design} with these criteria is a convex optimization problem. Indeed, it is known that \eqref{design} with
the above criteria can be reformulated as (possibly nonlinear) semidefinite programming (SDP) problems (see,
for example, \cite{FedLee00,BenNem01,BoyVan04,Papp10}).

The optimal design problems \eqref{design} with the aforementioned
criteria usually do not have closed form solutions. Numerous
procedures have thus been proposed to solve \eqref{design} (see, for example, \cite{Fed72,Wyn72,Atw73,Atw76,WuW78,Atw80,HaP07,Bohning86,Paz86,AtkinDonTob07,Tor07,AhSuTo08,Det08,Rich09,TMM09,Sag10}).
Among them, the multiplicative algorithm introduced in \cite{STT78}
has been widely explored. For example, Titterington \cite{Tit76},
P\'{a}zman \cite{Paz86}, Dette et al. \cite{Det08} and Harman and
Trnovsk\'{a} \cite{HaT09} studied the multiplicative algorithm for
D-criterion. In addition, Fellman \cite{Fel74} and Torsney
\cite{Tor83} considered the multiplicative algorithm for A-criterion
under the assumption that all $A_i$'s are rank-one. Recently, Yu
\cite{Yu10a} studied the multiplicative algorithm for a class of convex
optimality criteria and proved its global convergence under some
assumptions. Nevertheless, for several commonly used optimality criteria, some
of those assumptions may not hold and hence there is no theoretical
guarantee for its convergence. Indeed, as observed in \cite[Section
5]{Yu10a}, one of the assumptions does not hold for $p$th mean
criterion with $p = -2$. Moreover, for such a criterion, our
numerical experiments in Section \ref{sec:sim} demonstrate that the
multiplicative algorithm appears not to converge when $p < -1$. More details about
the multiplicative algorithm for solving \eqref{design} are given in
Section~\ref{sec:mult}.

In this paper, we consider an alternative approach to solve problem \eqref{design}.
In particular, we propose an interior point (IP) method for \eqref{design} and establish
its global convergence. The method is a Newton-type method that can be efficiently 
applied to solve problem \eqref{design} with a broad class of convex optimality 
criteria and moderate-sized matrices $A_i$'s. By exploiting the structure of the 
Hessian matrix of the classical A-, D- and $p$th mean criterion, we derive an 
explicit formula for its rank. Using this result, we further show that the Newton 
direction arising in the IP method for \eqref{design} with the aforementioned 
classical optimality criteria can be computed efficiently via 
Sherman-Morrison-Woodbury formula when $n \gg m^2$, i.e., when the 
size of $A_i$'s is small relative to the sample size.
We finally compare the IP method with the multiplicative algorithm.
The computational results show that the IP method
usually outperforms the multiplicative algorithm in both speed and
solution quality.

The rest of this paper is organized as follows. In
Subsection~\ref{sec:notation}, we introduce the notations that are
used throughout the paper. In Section~\ref{sec:mult}, we review the
multiplicative algorithm and address its convergence. In
Section~\ref{sec:IP}, we propose an IP method for
solving problem~\eqref{design} with a large class of convex optimality
criteria and address its convergence. In Section~\ref{sec:IP2},
we discuss how the IP method can be applied
to solve problem~\eqref{design} with criteria (i)--(iv) and
demonstrate how the Newton direction can be computed efficiently
when $n\gg m^2$. In
Section~\ref{sec:sim}, we conduct numerical experiments to test the
performance of the method and compare it with the
multiplicative algorithm. Finally,
we present some concluding remarks in Section~\ref{conclude}.

\subsection{Notations}\label{sec:notation}

In this paper, the symbol $\Re_{++}$ denotes the set of all positive
real numbers and $\Re^n$ denotes the $n$-dimensional Euclidean
space. For a vector $x\in\Re^n$ and $\cI\subseteq \{1,\ldots,n\}$,
$\|x\|$ denotes the Euclidean norm of $x$, $x_\cI$ denotes the
subvector of $x$ indexed by $\cI$ and $\D(x)$ denotes the diagonal
matrix whose $i$th diagonal entry is $x_i$ for all $i$. For $\alpha \in \Re$
and a vector $x\in \Re^n$ with positive entries, $x^{\alpha}$ denotes the
vector whose $i$th entry is $x_i^{\alpha}$ for all $i$. For $x$, $y\in \Re^n$,
$x\circ y$ denotes the Hadamard (entry-wise) product of $x$ and $y$. The letter $e$ denotes the
vector of all ones, whose dimension should be clear from the
context. The set of all $m \times n$ matrices with real entries is
denoted by $\Re^{m \times n}$. For any $A\in \Re^{m\times n}$,
$\cI\subseteq \{1,\ldots,m\}$ and $\cJ\subseteq \{1,\ldots,n\}$,
$a_{ij}$ denotes the $(i,j)$th entry of $A$, $A_\cJ$ denotes the
submatrix of $A$ comprising the columns of $A$ indexed by $\cJ$ and
$A_{\cI\cJ}$ denotes the submatrix of $A$ comprising the rows and
columns of $A$ indexed by $\cI$ and $\cJ$, respectively. The space
of $n \times n$ symmetric matrices will be denoted by $\cS^n$. If $A
\in \cS^n$ is positive semidefinite (resp., definite), we write $A
\succeq \ 0$ (resp., $A \succ 0$). The cone of positive semidefinite
(resp., definite) matrices is denoted by $\cS^n_+$ (resp.,
$\cS^n_{++}$). For $A,B\in \cS^n$, $A\succeq B$ (resp., $A \succ B$)
means $A-B \succeq 0$ (resp., $A-B \succ 0$).
The trace of a real square matrix $A$ is denoted by $\tr(A)$.
We denote by $I$ the identity matrix, whose dimension should be
clear from the context.


A function $f:\cS^n\rightarrow \Re$ is said to be increasing (resp., decreasing) if for any
$A\succeq B$, it holds that
\begin{equation*}
  f(A)\ge  f(B) \ \ (\mbox{resp.}, \ f(A) \le  f(B)).
\end{equation*}

\section{The multiplicative algorithm}\label{sec:mult}

In this section we review the multiplicative algorithm introduced in
\cite{STT78} for solving problem \eqref{design} and discuss its convergence.
In particular, we first describe the multiplicative algorithm as follows, which is
specified through a power parameter $\lambda\in (0,1]$.

\vgap

\noindent {\bf Multiplicative Algorithm:}
\begin{itemize}
   \item[1.] {\bf Start:} Let a positive $w^0 \in \Omega$ and $\lambda\in (0,1]$ be given.
  \item[2.] {\bf For} $k=0, 1, \ldots$
  \begin{equation}\label{power}w^{k+1}_i=w^k_i\frac{(d_i(w^k))^\lambda}{\sum_{j=1}^n
  w^k_j(d_j(w^k))^\lambda}, \quad i=1,\ldots, n,
  \end{equation}
  where $d_i(w)=-\tr(\nabla\Phi(\cM(w))A_i)$ and $\nabla
  \Phi(\cM(w))$ is the gradient of $\Phi$ at $\cM(w)$.\\
  {\bf End} (for)
\end{itemize}

\begin{remark}
The above algorithm is the same as the one described in \cite{Yu10a},
in the sense that both algorithms generate exactly the same
sequence $\{w^k\}$ provided the initial points $w^0$ are identical.
\end{remark}
\vgap

We now state a global convergence result recently established by Yu
\cite[Theorem~2]{Yu10a} for the multiplicative algorithm when applied to
solve the following problem, which is closely related to
\eqref{design}:
\begin{equation}\label{Yudesign}
\begin{array}{rl}
  {\rm val}:=\sup\limits_w & -\Phi(\cM(w))\\
  {\rm s.t.}& w\in \Omega,\ \cM(w)\succ 0.
\end{array}
\end{equation}
Observe that \eqref{design} and \eqref{Yudesign} are equivalent (i.e., the optimal value being negative of each other)
if there exists an optimal solution $w^*$ of \eqref{design} with $\cM(w^*)\succ 0$, or if $\Phi$ is convex in
$$\cS^m_{+}(K) := \{X \in \cS^m_{+}:\; \range(K) \subseteq \range(X)\}.$$
and \eqref{Yudesign} has an optimal solution.

\begin{proposition}\label{convergence_mult}
Let $\{w^k\}$ be the sequence generated from the above multiplicative algorithm.
Suppose the following assumptions hold:
  \begin{enumerate}[\rm (a)]
    \item for any feasible point $w$ of \eqref{Yudesign}, ${\nabla\Phi}(\cM(w))\preceq 0$
    and ${\nabla\Phi}(\cM(w))A_i\neq
    0$ for $i=1,\ldots,n$;
    \item for any feasible point $w$ of \eqref{Yudesign}, if $T(w)\neq
    w$, then $\Phi(\cM(T(w))) < \Phi(\cM(w))$,
    where \[[T(w)]_i \ := \ w_i\frac{(d_i(w))^\lambda}{\sum_{j=1}^n
  w_j(d_j(w))^\lambda}, \quad i=1,\ldots, n;\]
    \item $\Phi$ is strictly convex and $\nabla\Phi$ is continuous in
    $\cS^m_{++}$;
    \item for any $\{X^k\} \subset \cS^m_{++}$, if $X^k \to X^*$ and $\{\Phi(X^k)\}$ is
    decreasing, then $X^*\succ 0$.
  \end{enumerate}
  Then $\Phi(\cM(w^k))\rightarrow -{\rm val}$ monotonically, and
moreover, any accumulation point of $\{w^k\}$ is an optimal solution
of \eqref{Yudesign}.
\end{proposition}
\begin{remark}
  Notice that the assumptions in the above proposition imply that
  any accumulation point $w^*$ of $\{w^k\}$ satisfies $\cM(w^*)\succ 0$.
  Hence, if the assumptions in Proposition~\ref{convergence_mult} hold and $\Phi$ is convex in
$\cS^m_{+}(K)$, then \eqref{design}
  is equivalent to \eqref{Yudesign} and any accumulation point of the sequence $\{w^k\}$
  generated from the above multiplicative algorithm solves \eqref{design}.
\end{remark}
\vgap

Using Proposition~\ref{convergence_mult} and some technical results
developed in \cite{Yu10a}, one can establish the convergence of the
above multiplicative algorithm when applied to problem
\eqref{design} with A-, D- and $p$th mean criterion for $p\in
(-1,0)$ and $K=I$, which is summarized as follows.

\begin{corollary} \label{cor}
Assume that $K=I$ and $A_i \neq 0$ for $i=1,\ldots,n$. Then the multiplicative algorithm
converges for any $\lambda \in (0,1]$ when applied to problem \eqref{design} with D- and $p$th
mean criterion for $p\in (-1,0)$. Also, it converges for A-criterion when $\lambda \in (0,1)$.
\end{corollary}

As seen from Proposition~\ref{convergence_mult} and
Corollary~\ref{cor}, the multiplicative algorithm converges for a
large class of optimality criteria $\Phi$. Nevertheless, for some important
convex optimality criteria, the assumptions stated in
Proposition~\ref{convergence_mult} may not hold and hence there is
no theoretical guarantee for its convergence. Indeed, as observed in
\cite[Section 5]{Yu10a}, the assumption (b) with $\lambda=1$ does
not hold for $p$th mean criterion with $p = -2$. Moreover, for such
a criterion, our numerical experiments in Section \ref{sec:sim}
demonstrate that the multiplicative algorithm appears not to
converge when $p < -1$.

Due to the aforementioned potential drawbacks of the multiplicative algorithm, we will propose
an IP method for solving problem \eqref{design} with a broad class of optimality criteria $\Phi$
including A-, D- and $p$th mean criterion in subsequent sections.

\section{IP method for a class of convex optimality criteria} \label{sec:IP}

In this section, we propose an IP method for solving \eqref{design} with a class of convex
optimality criteria $\Phi = \Psi\circ {\cal C}_K$. We make the following assumption on $\Psi$
throughout this section.

\begin{assumption} \label{assump}
The function $\Psi$ is convex, decreasing, twice continuously
differentiable and bounded below in $\cS^m_{++}$. Moreover,
for any bounded sequences $\{X^k\}\subseteq \cS^m_{++}$ with $\lambda_{\min}(X^k)\rightarrow 0$,
one has $\Psi(X^k)\rightarrow \infty$.
\end{assumption}

\begin{remark}\label{remark3}
We now make some brief comments on the above assumptions.
\begin{enumerate}[{\rm (a)}]
\item Assumption~\ref{assump} is fairly reasonable. Indeed, all optimality criteria described in Section~\ref{intro}
satisfy this assumption.

  \item Since the feasible set is not necessarily closed, problem~\eqref{design} with a general convex optimality criterion
  may not have an optimal solution. However, when the optimality criterion satisfies Assumption~\ref{assump}, it must
have an optimal solution as shown in Theorem~\ref{convergence}(a). We refer the readers to
  \cite[Chapter~5]{Puk93} for more discussion on conditions guaranteeing
  existence of solutions for problem~\eqref{design}.
  \item In contrast to Proposition~\ref{convergence_mult}, we do not require the existence of a
  positive definite optimal moment matrix $\cM(w^*)$. Indeed, Assumption~\ref{assump} may hold
  even when problem~\eqref{design} does not have a positive definite optimal moment matrix.
  For instance, the design problem
  \begin{equation*}
    \begin{array}{rl}
      \min\limits_{w,X} & \begin{pmatrix}
        1\\0
      \end{pmatrix}^TX^\dagger \begin{pmatrix}
        1\\0
      \end{pmatrix}\\
      {\rm s.t.} & X = \begin{pmatrix}
        w_1&0\\0&w_2
      \end{pmatrix}, w_1+w_2=1, w_1,w_2\ge 0,\\
      &\begin{pmatrix}
        1\\0
      \end{pmatrix}\in {\rm Range}(X),
    \end{array}
  \end{equation*}
  has a unique optimal solution at $(w_1,w_2) = (1,0)$. The corresponding optimal moment matrix is not positive
  definite; thus, the assumption (d) of Proposition~\ref{convergence_mult} does not hold.
  However, it is easy to check that Assumption~\ref{assump} is satisfied for this design problem (with $\Psi(t)=1/t$).
  In general, the assumption (d) of Proposition~\ref{convergence_mult} is likely not satisfied when $K$ is not invertible,
  while our Assumption~\ref{assump} is independent of $K$.
\end{enumerate}
\end{remark}

Under Assumption~\ref{assump}, it is not hard to show that the function $\Phi(\cM(\cdot))$
is bounded below on the feasible set of \eqref{design}. Also, it is routine to show that the function $\Phi$
is twice continuously differentiable in $\cS^m_{++}$. Furthermore, it can be shown that $\Phi$ is convex
in $\cS^m_{+}(K)$ by considering suitable Schur complements (see, for example, \cite[Section~6]{Papp10}).
We include a short proof below for the convenience of readers. Before proceeding, we state the following
well-known fact, which concerns the Schur complement of a positive semidefinite submatrix (see, for example,
\cite[Lemma~3.12]{Puk93}).

\begin{lemma}\label{Schur2}
   Let $A\in \cS^k$, $B\in \Re^{m\times k}$ and $C\in \cS^m$. Then the matrix $\begin{pmatrix}
       A&B^T\\B&C
     \end{pmatrix}$ is positive semidefinite if and only if $A\succeq
     B^TC^{\dagger}B$, $C\succeq 0$ and ${\rm Range}(B)\subseteq {\rm Range}(C)$.
\end{lemma}

\begin{proposition}
The optimality criterion $\Phi$ is convex in $\cS^m_{+}(K)$.
\end{proposition}
\begin{proof}
First of all, it can be shown that the set $\cS^m_{+}(K)$ is convex (see, for example, \cite[Section~3.3]{Puk93}). In addition,
notice that for any $X\in \cS^m_{+}(K)$, we have
  \beqa
    \Phi(X)  &=& \Psi((K^TX^\dagger K)^{-1})  \ = \ \inf_U\left\{\Psi(U):\; (K^TX^\dagger K)^{-1}\succeq U\succ 0\right\} \nn \\
    & = &\inf_U\left\{\Psi(U):\; U^{-1}\succeq K^TX^\dagger K, U\succ 0\right\} \nn \\
    & = & \inf_U\left\{\Psi(U):\; \begin{pmatrix}
      U^{-1}&K^T\\K&X
    \end{pmatrix}\succeq 0, U\succ 0\right\} \nn \\
    & = & \inf_U\left\{\Psi(U):\; X\succeq KUK^T, U\succ 0\right\}, \label{phix}
  \eeqa
  where the second equality follows from the fact that $\Psi$ is decreasing, the
  fourth and last equalities follow from Lemma~\ref{Schur2}, while
  the third equality holds because $K^TX^\dagger K$ is invertible for $X\in \cS^m_{+}(K)$
  when $K$ has full column rank. Convexity of
  $\Phi$ in $\cS^m_{+}(K)$ now follows from \cite[Theorem~5.7]{Roc70}.
\end{proof}

\gap

Observe that $\cM(w)\succ 0$ whenever $w > 0$. Thus, under Assumption~\ref{assump},
the function $\Phi$ is twice continuously differentiable for any positive $w\in \Omega$.
It is hence natural to develop an IP method to solve \eqref{design}
since such a method keeps
all iterates in the relative interior of $\Omega$ until convergence.
To proceed, we first reformulate the problem by eliminating the
equality constraint. The resulting equivalent problem is given by
\begin{equation}\label{P0}
  \begin{array}{rl}
    f^* = \inf\limits_{\tw} &f(\tilde w):=\Phi(\cM(P\tilde w+q))\\
    {\rm s.t.}& e^T\tilde w\le 1, \ \tilde w\ge 0, \\
    & {\rm Range}(K)\subseteq {\rm Range}(\cM(P\tilde
    w+q)),
  \end{array}
\end{equation}
where $P\in \Re^{n\times (n-1)}$ and $q\in \Re^n$ are such that
\begin{equation}\label{Bh}
P\tilde w+q=\begin{pmatrix}
  \tw \\ 1-e^T\tw
\end{pmatrix}
 \quad \quad \forall \tw \in \Re^{n-1}.
\end{equation}

We next develop an IP method for solving problem \eqref{P0} instead.
First, we need to build a suitable barrier function.
Given any $\tw > 0$ satisfying $e^T \tw < 1$, one can observe that
$P\tw+q > 0$ and hence $\cM(P\tw+q) \succ 0$, which leads to
${\rm Range}(K)\subseteq {\rm Range}(\cM(P\tilde w+q))$. This implies that any
barrier function that takes into account the first two inequality
constraints of \eqref{P0} is sufficient for the development of IP
method. Here we naturally choose the logarithmic barrier function
and then solve the barrier subproblem in the form of
\begin{equation}\label{minfmu}
\min_{\tilde w} f_\mu(\tilde w):=f(\tilde w)-\mu
\sum_{i=1}^{n-1}\log(\tilde w_i)-\mu \log\left(1-e^T\tilde w\right)
\end{equation}
for a sequence of parameters $\mu \downarrow 0$. In view of
Assumption \ref{assump}, we see that any level set of
$f_\mu$ is compact. Moreover, $f_\mu$ is strictly convex. Thus,
there exists a unique minimizer to \eqref{minfmu} for any $\mu >0$.
Furthermore, it follows from Assumption \ref{assump} that $f_\mu$ is
twice continuously differentiable and its Hessian is positive
definite in its domain. Therefore, problem \eqref{minfmu} can be
suitably solved by the Newton's method with a line search whose
stepsize is chosen by Armijo rule.

We are now ready to present our IP method for solving problem
\eqref{P0}.

\vgap

\noindent{\bf IP Method:}
\begin{itemize}
  \item[1.] {\bf Start:} Let a strictly feasible $\tw^0$, $0<\beta,\gamma,\eta,\sigma<1$ and $\mu_1>0$ be given.
  Let $\epsilon(\mu)$ be an increasing function of $\mu$ so
  that $\lim_{\mu\downarrow 0}\epsilon(\mu)=0$. Set $\tw=\tw^{0}$ and $k=1$.
  \item[2.] {\bf While} $\|\nabla f_{\mu_k}(\tilde w)\| > \epsilon(\mu_k)$ {\bf
  do}
  \begin{enumerate}[(a)]
    \item Compute the Newton direction
  \begin{equation}\label{Ndirection}
  d:=-(\nabla^2f_{\mu_k}(\tilde w))^{-1}\nabla f_{\mu_k}(\tilde w).
  \end{equation}
    \item Let $\alpha_{\max}(\tilde w):=\max\{\alpha:\;\tilde w[\alpha] \ge 0, \ e^T\tilde
  w[\alpha] \le 1\}$, where $\tilde w[\alpha]:=\tilde w+\alpha d$.
    \item Let $\alpha$ be the largest element of $\{\balpha(\tilde w),\beta\balpha(\tilde w),\beta^2\balpha(\tilde w),\cdots\}$ satisfying
  \begin{equation*}
    f_{\mu_k}(\tilde w[\alpha])\le f_{\mu_k}(\tilde w)+ \sigma\alpha (\nabla
    f_{\mu_k}(\tilde w))^Td,
  \end{equation*}
  where $\balpha(\tilde w) := \min\{1,\eta\alpha_{\max}(\tilde w)\}$.
    \item Set  $\tilde w \leftarrow \tilde w[\alpha]$.
  \end{enumerate}
  {\bf End} (while)
  \item[3.] Set $\tilde w^{k}\leftarrow \tw$, $\mu_{k+1} \leftarrow \gamma \mu_k$, $k\leftarrow k+1$, and
   go to step 2.
\end{itemize}



In standard convergence analysis of IP methods,
the feasible sets are usually assumed to be closed and the
objective functions are twice continuously differentiable in a
neighborhood of the feasible sets (see, for example, \cite{ForGilWri02}).
Nevertheless, these two conditions do not necessarily hold for our problem \eqref{P0}.
In particular, the objective function is not necessarily continuous
up to the boundary of the feasible region \cite[Section~3.16]{Puk93}.
Hence, it is not immediately clear the sequence generated by our method
will accumulate at a global minimizer of \eqref{P0}. Thus,
we discuss convergence of our IP method below.
We first present convergence results regarding the outer iterations of our IP
method and then discuss the convergence of its inner
iterations.

For notational convenience, in the remainder of this section,
we associate with each $\tw\in \Re^{n-1}$ a unique $w\in\Re^n$ by
letting $w := P\tw+q $. Analogously, we associate with each $w\in \Re^n$
a unique $\tw\in\Re^{n-1}$ by letting $\tw_i = w_i$ for $i=1,\ldots,n-1$.
Also, we let $\Phi_\cM(w) := \Phi(\cM(w))$.

 We first observe that if problem \eqref{design} has an optimal solution
$w^*$ with $\cM(w^*)\succ 0$, 
 then there exists a Lagrange multiplier $u^*\ge 0$
such that $(w^*,u^*)$ satisfies the following KKT system:
\begin{align}\label{KKToriginal}
\begin{array}{rl}
P^T(\nabla\Phi_\cM(w)-u)&=\ 0,\\
  e^Tw&=\ 1,\\
  u\circ w&=\ 0,\\
  (w, u)&\ge\  0.
\end{array}
\end{align}
Given a strictly feasible point $\tw\in \Re^{n-1}$ of problem
\eqref{minfmu}, we notice that
\begin{equation} \label{grad}
\nabla f_\mu(\tilde w)=P^T(\nabla \Phi_\cM(w)-\mu w^{-1}).
\end{equation}
Then it is not hard to observe that for each $\mu >0$, the $w$
associated with the approximate solution $\tw$ of \eqref{minfmu}
obtained by the Newton's method detailed in step 2 above together
with $u:=\mu w^{-1}$ satisfies the following perturbed KKT system:
\begin{align}\label{KKTfmuapprox}
\begin{array}{rl}
P^T(\nabla\Phi_\cM(w)-u)&=\ v,\\
  e^Tw&=\ 1,\\
  u\circ w&=\ \mu e,\\
  (w,u) &>\ 0
\end{array}
\end{align}
for some $v\in \Re^{n-1}$. The convergence regarding the outer iterations of
our IP method is related to the limiting behavior of the
solutions of system \eqref{KKTfmuapprox} as $(\mu,v) \rightarrow
(0_+,0)$, that is, $(\mu,v) \rightarrow (0,0)$ with $\mu>0$.

We first claim that system \eqref{KKTfmuapprox} has a unique solution for
any $(\mu, v)\in \Re_{++} \times \Re^{n-1}$. Indeed, it is easy to observe
that $(w,u)$ is a solution of \eqref{KKTfmuapprox} if and only if $\tw \in \Re^{n-1}$
is an optimal solution of
\beq \label{KKT-p}
\min\limits_\tw f_\mu(\tilde w)-v^T\tilde w.
\eeq
Since the objective function of
\eqref{KKT-p} is strictly convex and it has compact level sets,
problem \eqref{KKT-p} has a unique optimal solution, which immediately
implies that system \eqref{KKTfmuapprox} has a unique solution. From now on,
we denote by $(w(\mu,v),u(\mu,v))$ the unique solution of \eqref{KKTfmuapprox}.
Our main theorem below discusses the limiting behavior of $(w(\mu,v),u(\mu,v))$ as $(\mu, v)\rightarrow (0_+,0)$.
The proof of this theorem can be found in the appendix.

\begin{theorem}\label{convergence}
Let $(w(\mu,v),u(\mu,v))$ be defined above for $(\mu, v)\in \Re_{++} \times \Re^{n-1}$.
Then the following statements hold:
  \begin{enumerate}[{\rm (a)}]
    \item $\lim\limits_{(\mu,v) \rightarrow (0_+,0)} \Phi(\cM( w(\mu,v))) =
    f^*$ and any accumulation point of $w(\mu,v)$ as $(\mu,v) \rightarrow (0_+,0)$ is an optimal solution of \eqref{design}.
    \item Suppose in addition that problem \eqref{design} has an optimal solution $w^*$ with $\cM(w^*)\succ 0$. Then any
    accumulation point of $w(\mu,v)$ as $(\mu,v) \xrightarrow[\Xi_C]{} (0,0)$,
    i.e., $(\mu,v) \rightarrow (0,0)$ with $(\mu,v)\in \Xi_C:=\{(\mu,v):\;\|v\|_\infty<C\mu\}$
    for some given $C>0$, is an optimal solution of \eqref{design} with maximum cardinality.
  \end{enumerate}
 \end{theorem}

\vgap

As an immediate consequence of Theorem~\ref{convergence}, we have the following
global convergence result regarding the outer iterations
of our IP method, whose simple proof is omitted.

\begin{corollary} \label{cor-1}
Let $\{\mu_k\}$ and $\{\tw^k\}$ be the sequences generated in the
 IP method. Let $w^k=P\tw^k+q$ for all $k$. Then the following
statements hold:
  \begin{enumerate}[{\rm (a)}]
    \item $\lim\limits_{k\rightarrow \infty} \Phi(\cM(w^k)) = f^*$
    and any accumulation point of $\{w^k\}$ is an optimal solution of \eqref{design}.
    \item Suppose in addition that problem \eqref{design} has an optimal solution $w^*$ with $\cM(w^*)\succ 0$ and $\epsilon(\mu_k)=O(\mu_k)$. Then any
    accumulation point of $\{w^k\}$ is an optimal solution of \eqref{design} with maximum cardinality.
  \end{enumerate}
\end{corollary}
%


We emphasize that in Corollary~\ref{cor-1} (a), we do {\em not} require
existence of an optimal solution $w^*$ with $\cM(w^*)\succ 0$. On the other hand, if
such an optimal solution does exist,
for example, when $K = I$, then Corollary~\ref{cor-1} (b) states that the accumulation point (with $\epsilon(\mu_k)=O(\mu_k)$)
must be an optimal solution of \eqref{design} that has the largest number of non-zero entries among all
the optimal solutions of \eqref{design}.

Before ending this section, we establish a convergence result regarding the inner
iterations of our IP method.

\begin{proposition}
  Let $\mu_k >0$ and $\epsilon(\mu_k)>0$ be given. Then the Newton's method detailed in step 2 of
  the  IP method starting from any strictly feasible point $\tw^{_{\rm init}}$ of \eqref{P0}
  generates a point $\tw^k$ satisfying $\|\nabla f_{\mu_k}(\tw^k)\|\le \epsilon(\mu_k)$ within a finite number of iterations.
\end{proposition}
\begin{proof}
First, observe that all iterates generated by the Newton's
method lie in the compact level set $\Upsilon
:=\{\tw:\;f_{\mu_k}(\tilde w)\le f_{\mu_k}(\tw^{_{\rm init}})\}$.
Furthermore, it holds that $\tw >0$ and $1-e^T\tw > 0$ for all $\tw \in \Upsilon$. This together with the
assumption that $\cM(\Omega)\cap \cS^m_{++}\neq \emptyset$ implies that
$\cM(\Upsilon) \subset \cS^m_{++}$. Thus $\nabla f_{\mu_k}$ and
$\nabla^2 f_{\mu_k}$ are continuous in $\Upsilon$. Using this
observation and the strong convexity of $f_{\mu_k}$ in $\Upsilon$,
there exist $\underline{\lambda}$, $\overline{\lambda}>0$ such that
  $\underline{\lambda} I\preceq \nabla^2f_{\mu_k}(\tw)\preceq \overline{\lambda} I$ for
  all $\tw \in \Upsilon$.
This relation along with the continuity of $\nabla f_{\mu_k}$ and
$\nabla^2 f_{\mu_k}$ implies that $d =
-(\nabla^2f_{\mu_k}(\tw))^{-1}\nabla f_{\mu_k}(\tw)$ is continuous
in $\Upsilon$. In view of this result and the definition of
$\balpha(\tw)$, it is not hard to show that $\balpha(\tw)$ is
positive and continuous in $\Upsilon$. This fact together with the
compactness of $\Upsilon $ yields
$\underline{\alpha}:=\inf\{\balpha(\tilde w):\;\tw \in \Upsilon
\}>0$. Thus, all iterates $\tw$ generated by the Newton's method
satisfy $\underline{\lambda} I\preceq \nabla^2f_{\mu_k}(\tw)\preceq
\overline{\lambda}  I$ and $\balpha(\tw)\in[\underline{\alpha},1]$.
The remaining proof follows the same arguments as in the proof of
\cite[Theorem~3.13]{LuZ09}.
\end{proof}

\section{IP method for classical optimality criteria} \label{sec:IP2}

In this section, we discuss how to apply our IP method to solve problem~\eqref{design}
with A-, D- and $p$th mean criterion. In particular, we will demonstrate how the Newton
direction \eqref{Ndirection} can be efficiently computed for each criterion.

Before proceeding, we introduce some notations that will be used in this section (see,
for example, \cite{TTT98} for more details).  Given matrices $A$ and $B$ in
$\Re^{m \times n}$, $A\otimes B$ denotes the Kronecker
product of $A$ and $B$, while $A\circ B$ denotes the Hadamard (entry-wise) product of
$A$ and $B$. In addition, $\vec(A)$ denotes the column vector formed by stacking
columns of $A$ one by one. For any $m \times m$ symmetric matrix $U$, we define the vectors
$\svec(U) \in \Re^{m(m+1)/2}$ and $\svecr(U) \in \Re^{m(m+1)/2}$ as
\[
\begin{split}
\svec(U) &= (u_{11},\sqrt2 u_{21},\ldots,\sqrt2 u_{m1},u_{22},\sqrt2 u_{32},\ldots,\sqrt2 u_{m2},
\ldots, u_{mm})^T.\\
\svecr(U) &= (u_{11},u_{21},\ldots,u_{m1},u_{22},u_{32},\ldots,u_{m2},
\ldots, u_{mm})^T.
\end{split}
\]
It is not hard to observe that $\svec$ is an isometry between $\cS^m$ and $\Re^{m(m+1)/2}$ and
moreover,
\beq \label{UV}
\tr(UV) = \svec(U)^T\svec(V) \ \ \ \forall U, V \in \cS^m.
\eeq
We denote the inverse map of $\svec$ by $\smat$. Clearly, they are adjoint of each other, namely,
\[
u^T \svec(V) = \tr(\smat(u)V) \ \ \ \forall u \in \Re^{m(m+1)/2}, V \in \cS^m.
\]

The symmetric Kronecker product of any two (not necessarily  symmetric) matrices
$G, H \in \Re^{m\times m}$ is a square matrix of order $m(m+1)/2$ such that
\beq \label{sotimes}
(G \sotimes H) \svec(U) = \frac12\svec(GUH^T+HUG^T) \ \ \ \forall U \in \cS^m.
\eeq
As mentioned in \cite{TTT98}, $G \sotimes H$ can be expressed in terms of the standard
Kronecker product of $G$ and $H$ as follows:
\[
G \sotimes H  = \frac12 Q(G\otimes H + H\otimes G)Q^T,
\]
where $Q \in \Re^{m(m+1)/2 \times m^2}$ is such that
\beq \label{Q}
Q\vec(U) = \svec(U), \ \ \ Q^T \svec(U) = \vec(U) \ \ \ \forall U \in \cS^m.
\eeq
It is easy to observe that the above $Q$ exists and is unique. Moreover, $QQ^T=I$.

Throughout this section, for each optimality criterion $\Phi$, we define the associated
function $\phi$ as follows:
\beq \label{phi}
\phi(x) = \Phi(\smat(x))
\eeq
for any $x\in \Re^{m(m+1)/2}$, provided that $\Phi(\smat(x))$ is well-defined. It is clear to
observe that $\phi$ is convex due to the convexity of $\Phi$. Define
\[
M := [\svec(A_1) \dots \svec(A_n)].
\]
Clearly, $M \in \Re^{m(m+1)/2 \times n}$.

With the notations above, the function $f_{\mu}$ defined in \eqref{minfmu} can be rewritten as
\[
f_{\mu}(\tw) = \phi(M (P\tw+q)) - \mu
\sum_{i=1}^{n-1}\log(\tilde w_i)-\mu \log\left(1-e^T\tilde w\right).
\]
By the chain rule, the gradient and Hessian of $f_\mu$ are given by
\begin{align}
  \nabla f_\mu(\tw) & = P^TM^T\nabla \phi(Mw)-\mu P^Tw^{-1},\nonumber\\
  \nabla^2 f_\mu(\tw)
  & = P^TM^T\nabla^2 \phi(Mw)MP + \frac{\mu}{(1-e^T\tilde w)^2}ee^T + \mu \D(\tilde w^{-2}),\label{p_hessian}
\end{align}
where $w=P\tw+q$.

The main computational effort of our IP method lies in computing the Newton direction $d$ by solving the system
$\nabla^2 f_\mu(\tilde w)d =-\nabla f_\mu(\tilde w)$ (see \eqref{Ndirection}).
In applications, $n$ can be significantly larger than $m^2$. Since the rank of $\nabla^2 \phi(Mw)$ is at most
$m(m+1)/2$, the first matrix in \eqref{p_hessian} has ``low'' rank compared to $\nabla^2 f_\mu(\tw)$.
It is generally more efficient to compute the Newton direction via the Sherman-Morrison-Woodbury formula,
{\em without} explicitly forming the Hessian matrix.
To this end, suppose that $\nabla^2 \phi(Mw)$ has rank $r$. Let $VDV^T$ be the {\it partial} eigenvalue decomposition of $\nabla^2 \phi(Mw)$,
where $D$ is the $r \times r$ diagonal matrix whose diagonal consists of $r$ largest eigenvalues of $\nabla^2 \phi(Mw)$, and
the columns of $V$ are the corresponding eigenvectors.\footnote{The partial eigenvalue decomposition can be efficiently computed by the package PROPACK \cite{Lar}.}  Due to the convexity of $\phi$, one can observe that $\nabla^2 \phi(Mw)=VDV^T$.
It then follows from \eqref{p_hessian} that
\begin{equation*}
\begin{split}
\nabla^2 f_\mu(\tw) &= (P^TM^TV)D(V^TMP) + \frac{\mu}{(1-e^T\tilde w)^2}ee^T + \mu \D(\tilde w^{-2})\\
&= \begin{pmatrix}
  P^TM^TV & e
\end{pmatrix}\begin{pmatrix}
  D&0\\0& \frac{\mu}{(1-e^T\tilde w)^2}
\end{pmatrix}\begin{pmatrix}
  V^TMP \\ e^T
\end{pmatrix} + \mu \D(\tilde w^{-2}),
\end{split}
\end{equation*}
which together with the Sherman-Morrison-Woodbury formula yields the Newton direction
\[
d = -\left(\nabla^2 f_\mu(\tilde w)\right)^{-1} \nabla f_\mu(\tilde w) =
-\left[\frac{1}{\mu}\D(\tilde w^2) - \frac{1}{\mu^2}\D(\tilde w^2)
\begin{pmatrix}
    P^TM^TV & e
  \end{pmatrix}
  W
  \begin{pmatrix}
    V^TMP\\ e^T
  \end{pmatrix}\D(\tilde w^2)\right]\nabla f_\mu(\tilde w),
\]
where
\[
W = \left(\begin{pmatrix}
  D^{-1} & 0\\
  0 & \frac{(1-e^T\tilde w)^2}{\mu}
  \end{pmatrix} + \frac{1}{\mu}\begin{pmatrix}
    V^TMP\\ e^T
  \end{pmatrix}\D(\tilde w^2)\begin{pmatrix}
    P^TM^TV& e
  \end{pmatrix}\right)^{-1}.
\]
When $n \gg m^2$, the above approach is much more efficient than solving the Newton system directly by
performing Cholesky factorization of $\nabla^2 f_\mu(\tilde w)$. We remark that
the ideas of using Sherman-Morrison-Woodbury formula to solve specially structured Newton systems
have been explored in literature (see, for example, \cite{AndRooTerTraWar00,FerMun03}).

As seen from above, $\nabla \phi(Mw)$ and $\nabla^2 \phi(Mw)$ are needed to compute Newton direction.
Furthermore, since the Hessian
tends to become more ill-conditioned as $\mu\rightarrow 0$,
it is more desirable to explicitly determine the rank $r$ of $\nabla^2 \phi(Mw)$ a priori than to
use the numerical rank obtained from the Matlab built-in function in each iteration.
For the rest of this section, we will discuss how to evaluate $\nabla \phi(Mw)$ and $\nabla^2 \phi(Mw)$ for A-, D- and
$p$th mean criterion, and determine the rank $r$ of $\nabla^2 \phi(Mw)$ used in the
aforementioned partial eigenvalue decomposition of $\nabla^2 \phi(Mw)$. The latter quantity turns out to be
independent of $w > 0$.

\subsection{IP method for $p$th mean criterion}
\label{sec:p}

Recall from Section~\ref{intro} that in $\cS^m_{++}$, the $p$th mean criterion $\Phi$ becomes
\begin{equation}\label{IP2-p-mean}
\Phi(X) = \tr((K^TX^{-1}K)^{-p})
\end{equation}
for some $p<0$ and $K\in \Re^{m\times k}$ with full column rank. It is easy to check that
Assumption~\ref{assump} holds for $\Phi$. Hence, problem \eqref{design} with this criterion
can be suitably solved by our IP method proposed in Section~\ref{sec:IP}.


Based on the above discussion, we know that our IP method needs the gradient and Hessian of
the associated function $\phi$ for computing Newton direction, where $\phi$ is defined by
\eqref{phi}. We next discuss how to compute them. Before proceeding, we state the following
classical result (see, for example, \cite[Proposition~4.3]{CQT03}) that will be used subsequently.

\begin{lemma} \label{deriv-g}
Let $g:\Re\rightarrow\Re$ be a differentiable function
and let $g^\square:\cS^m\rightarrow \cS^m$ be
defined by
\[ g^\square(Y):=V \begin{pmatrix}
  g(d_1)&&&\\
  &g(d_2)&&\\
  &&\ddots&\\
  &&&g(d_m)
\end{pmatrix}V^T,\]
where $V\D(d)V^T$ is an eigenvalue decomposition of $Y$ for some $d\in\Re^m$. Then the function $g^\square$
is well-defined, i.e., it is independent of the choice of $V$ and $d$,
and is also differentiable. Moreover, let
$S^{g,d}\in \cS^m$ be a symmetric matrix whose $(i,j)$th entry is given by
\begin{equation*}
s^{g,d}_{ij}:=\begin{cases}
  \displaystyle\frac{g(d_i)-g(d_j)}{d_i-d_j} &{\rm if}\ d_i\neq d_j, \\
  g'(d_i) & {\rm otherwise}.
\end{cases}\end{equation*}
Then the directional derivative
of $g^\square$ at $Y$ along the direction $H\in \cS^m$ is given by
\begin{equation*}
  V(S^{g,d}\circ (V^THV))V^T.
\end{equation*}
\end{lemma}

\begin{proposition}\label{grad-Hess}
  Let $\Phi$ be defined in \eqref{IP2-p-mean} and the associated $\phi$ be defined in \eqref{phi}.
  Let  $Q\in \Re^{m(m+1)/2\times m^2}$ be defined in \eqref{Q}. Then
the gradient and Hessian of $\phi$ at any $x \in \svec(\cS^m_{++})$ are given by
 \begin{align}
    \nabla\phi(x)&= p\svec(X^{-1}K(K^TX^{-1}K)^{-p-1}K^TX^{-1}),\label{Kgrad}\\
    \nabla^2\phi(x)&=Q(-p[(X^{-1}KV)\otimes (X^{-1}KV)] \D(\vec(S^{g,d}))
   [(X^{-1}KV)\otimes (X^{-1}KV)]^T\nonumber\\&\quad -p\ X^{-1}\otimes G-p\ G \otimes X^{-1})Q^T,\label{KHess}
  \end{align}
  respectively, where $X=\smat(x)$, $V\D(d)V^T$ is an eigenvalue decomposition of $K^TX^{-1}K$ for some $d\in \Re^m$,
  $g(t)=t^{-p-1}$, and $G=X^{-1}K[K^TX^{-1}K]^{-p-1}K^TX^{-1}$.
  In particular, when $K=I$, the above gradient and Hessian
  reduce to
  \begin{align}
    \nabla\phi(x)&= p \svec(X^{p-1}),\label{Igrad}\\
    \nabla^2\phi(x)&=(V\sotimes V) \D(\svecr(S^{g,d})) (V\sotimes V)^T,\label{IHess}
  \end{align}
  where $g(t)=pt^{p-1}$ and $V\D(d)V^T$ is an eigenvalue decomposition of $X$ for some $d\in \Re^m$.
\end{proposition}

\begin{proof}
To derive the gradient of $\phi$, we fix an arbitrary $x \in \svec(\cS^m_{++})$. Let $X=\smat(x)$.
For all sufficiently small $h\in \Re^{m(m+1)/2}$, we have $X+H \succ 0$, where $H=\smat(h)$, and moreover,
\beq \label{X-inv}
(X+H)^{-1} = X^{-1} - X^{-1}HX^{-1} + o(H).
\eeq
Using \eqref{X-inv} and Lemma \ref{deriv-g} with $g(t) = t^{-p}$ and $Y=K^TX^{-1}K$, we obtain that
\begin{align}\label{grad1}
  \Phi(X+H) &=
  \tr((K^T[X+H]^{-1}K)^{-p}) = {\rm
  tr}((K^TX^{-1}K-K^TX^{-1}HX^{-1}K+o(H))^{-p})\nonumber\\
  & = \Phi(X) -{\rm
  tr}(V(S^{g,d}\circ (V^TK^TX^{-1}HX^{-1}KV))V^T)+o(H),
\end{align}
where $V\D(d)V^T$ is an eigenvalue decomposition of $Y$. Letting $R:=-K^TX^{-1}HX^{-1}K$ and using the
fact that $V^T V=I$ and $s^{g,d}_{ii} = -pd_i^{-p-1}$ for all $i$, we further have
\begin{align}\label{grad2}
  \tr(V(S^{g,d}\circ (V^TRV))V^T) &= \tr(S^{g,d}\circ (V^TRV))
  =\sum_{i=1}^ms^{g,d}_{ii}\sum_{j,k}v_{ji}r_{jk}v_{ki}\nonumber\\
  & = -p
  \sum_{j,k}\left(\sum_{i=1}^mv_{ji}
  d_i^{-p-1}v_{ki}\right)r_{jk}= -\tr(p(K^TX^{-1}K)^{-p-1}R)\nonumber\\
  &= \tr(pX^{-1}K(K^TX^{-1}K)^{-p-1}K^TX^{-1}H).
\end{align}
In view of the definitions of $\phi$, $\Phi$, $X$ and $H$, it follows from \eqref{grad1}, \eqref{grad2} and
\eqref{UV} that
\[
\phi(x+h) - \phi(x) = \Phi(X+H) - \Phi(X) = h^T\left(p\svec(X^{-1}K(K^TX^{-1}K)^{-p-1}K^TX^{-1})\right) + o(h),
\]
which yields \eqref{Kgrad}.  And \eqref{Igrad} immediately follows from \eqref{Kgrad} by letting $K=I$.

We next derive the Hessian of $\phi$ at any $x \in \svec(\cS^m_{++})$. To proceed, we first recall the
following well-known results (see, for example, page~243 and Lemma~4.3.1 of \cite{HJ08}):
 \begin{align}\label{vecformula}
   \vec(ABC) = (C^T\otimes A)\vec(B),\qquad (A\otimes B)^T = A^T\otimes B^T.
\end{align}
Let $X$, $h$ and $H$ be defined as above. Using \eqref{X-inv} and Lemma \ref{deriv-g} with
$g(t) = t^{-p-1}$ and $Y=K^TX^{-1}K$, we have
\begin{align}\label{Hess1}
  \nabla \Phi(X+H) & = p(X+H)^{-1}K[K^T(X+H)^{-1}K]^{-p-1}K^T(X+H)^{-1}\nonumber\\
  &=p(X^{-1}-X^{-1}HX^{-1})K[K^T(X^{-1}-X^{-1}HX^{-1})K]^{-p-1}K^T(X^{-1}-X^{-1}HX^{-1})+o(H)\nonumber\\
  &=
  \nabla \Phi(X)-p(X^{-1}K)V(S^{g,d}\circ (V^TK^TX^{-1}HX^{-1}KV))V^T(K^TX^{-1})\nonumber\\
&\ \ \; -p GHX^{-1}-pX^{-1}HG+o(H),
\end{align}
where $G$ is defined as above. Since $X$ is symmetric, it follows from \eqref{vecformula} that
\begin{align}\label{Hess2}
  &\vec((X^{-1}KV)(S^{g,d}\circ
  (V^TK^TX^{-1}HX^{-1}KV))(V^TK^TX^{-1}))\nonumber\\
= \ &[(X^{-1}KV)\otimes (X^{-1}KV)]\vec(S^{g,d}\circ
  (V^TK^TX^{-1}HX^{-1}KV))\nonumber\\
= \ &[(X^{-1}KV)\otimes (X^{-1}KV)] \D(\vec(S^{g,d}))\vec(V^TK^TX^{-1}HX^{-1}KV))\nonumber\\
= \ &[(X^{-1}KV)\otimes (X^{-1}KV)] \D(\vec(S^{g,d})) [(X^{-1}KV)\otimes (X^{-1}KV)]^T \vec(H).
\end{align}
In addition, since $G$ is symmetric, we further have that
\begin{align}\label{Hess3}
\vec(GHX^{-1}+ X^{-1}HG)
 = \left[X^{-1}\otimes G + G \otimes X^{-1}\right]\vec(H).
\end{align}
In addition, by virtue of \eqref{Q}, \eqref{phi}, the definition of $X$ and $H$, and the fact
that $\svec$ is the adjoint operator of $\smat$, one can have
\[
\nabla \phi(x+h) - \nabla \phi(x) = \svec(\nabla\Phi(X+H)-\nabla\Phi(X))
= Q \vec(\nabla\Phi(X+H)-\nabla\Phi(X)).
\]
This relation together with \eqref{Q}, \eqref{Hess1}--\eqref{Hess3} and the definition of $H$ yields
\[
\ba{lcl}
\nabla \phi(x+h) - \nabla \phi(x) &=& Q(-p[(X^{-1}KV)\otimes (X^{-1}KV)] \D(\vec(S^{g,d}))
   [(X^{-1}KV)\otimes (X^{-1}KV)]^T \\
& & -p\ X^{-1}\otimes G-p\ G \otimes X^{-1})\vec(H) + o(Q\vec(H)) \\ [5pt]
&=& Q(-p[(X^{-1}KV)\otimes (X^{-1}KV)] \D(\vec(S^{g,d}))
   [(X^{-1}KV)\otimes (X^{-1}KV)]^T \\
& & -p\ X^{-1}\otimes G-p\ G \otimes X^{-1})Q^T\svec(H) + o(\svec(H)) \\ [5pt]
&=& Q(-p[(X^{-1}KV)\otimes (X^{-1}KV)] \D(\vec(S^{g,d}))
   [(X^{-1}KV)\otimes (X^{-1}KV)]^T \\ [5pt]
& & -p\ X^{-1}\otimes G-p\ G \otimes X^{-1})Q^T h + o(h),
\ea
\]
and hence \eqref{KHess} holds.

For the case when $K=I$, $\nabla^2\phi$ can be directly derived as follows. We know from
\eqref{Igrad} that $\nabla \Phi(X) = pX^{p-1}$. Letting $g(t) = p\,t^{p-1}$ and $V\D(d)V^T$ be
an eigenvalue decomposition of $X$, it follows
from Lemma \ref{deriv-g} that
\[
\nabla \Phi(X+H) = \nabla \Phi(X) + V(S^{g,d}\circ (V^THV))V^T + o(H).
\]
In view of \eqref{sotimes}, one can see that
\begin{align*}
  \svec(V(S^{g,d}\circ (V^THV))V^T)&=(V\sotimes V)\svec(S^{g,d}\circ
  (V^THV))\nonumber\\
  &=(V\sotimes V)[\svecr(S^{g,d})\circ \svec(V^THV)]\nonumber\\
  &=(V\sotimes V)(\svecr(S^{g,d})\circ [(V\sotimes V)^T\svec(H)])\nonumber\\
  &=(V\sotimes V) \D(\svecr({S^{g,d}})) (V\sotimes V)^T\svec(H).
\end{align*}
Using these relations and a similar proof as above, we can see that
\eqref{IHess} holds.
\end{proof}

\vgap

As mentioned earlier, we need to know the rank of $\nabla^2 \phi(x)$ for performing the partial eigenvalue
decomposition of $\nabla^2 \phi(x)$ which is used to compute Newton direction. In the next proposition, we
determine the rank of $\nabla^2 \phi(x)$ at any $x\in \svec(\cS^m_{++})$.

\begin{proposition}\label{rank_p}
Let $\Phi$ be defined in \eqref{IP2-p-mean} and the associated $\phi$ be defined in \eqref{phi}.
Then the rank of $\nabla^2 \phi(x)$ is $m(m+1)/2 - (m-k)(m-k+1)/2$ for any $x\in \svec(\cS^m_{++})$.
\end{proposition}

\begin{proof}
Let $x\in \svec(\cS^m_{++})$ be arbitrarily chosen. Define $X=\smat(x)$. Let $G$, $V$, $d$ and
$S^{g,d}$ be defined in Proposition~\ref{grad-Hess} with $g(t)=t^{-p-1}$. For convenience, we define
\[
\ba{lcl}
M_1 &=& [(X^{-1}KV)\otimes (X^{-1}KV)] \D(\vec(S^{g,d}))
  [(X^{-1}KV)\otimes (X^{-1}KV)]^T, \\ [4pt]
M_2 &=& X^{-1}\otimes G + G \otimes X^{-1}.
\ea
\]
To determine the rank of $\nabla^2 \phi(x)$, it suffices to know the dimension of the null space
of $\nabla^2 \phi(x)$, denoted by $\Null(\nabla^2\phi(x))$. Notice that $\phi$ is a twice
differentiable convex function in $\svec(\cS^m_{++})$. Thus, $\nabla^2 \phi(x) \succeq 0$. It
implies that $h \in \Null(\nabla^2\phi(x))$ if and only if $h^T\nabla^2\phi(x)h=0$. We will subsequently
show that
\beq \label{null-space}
h^T\nabla^2\phi(x)h=0 \ \Leftrightarrow \ K^TX^{-1}H = 0,
\eeq
where $H=\smat(h)$.
It then follows that
\[
h \in \Null(\nabla^2\phi(x)) \Leftrightarrow \ K^TX^{-1}H = 0.
\]
Notice that $K^TX^{-1}$ has full row rank. Thus, there exist nonsingular
matrices $E_1$ and $E_2$ such that
    $K^TX^{-1} = E_1\begin{pmatrix}
      I & 0\end{pmatrix}E_2$,
  where $I$ is the identity matrix of order $k$. It then follows that
\[
K^TX^{-1}H = 0 \ \Leftrightarrow \
      \begin{pmatrix}
      I & 0
    \end{pmatrix} U = 0,
  \]
  where $U = E_2HE_2^T\in \cS^m$. It is easy to see that the dimension of $\{U\in\cS^m:\; \begin{pmatrix}
      I & 0
    \end{pmatrix} U = 0\}$
is $(m-k)(m-k+1)/2$.
  Since $E_2$ is invertible, we conclude that the dimension of $\{H\in\cS^m:\;K^TX^{-1}H = 0\}$
is also $(m-k)(m-k+1)/2$. Since $\smat$ is a one-to-one map between $\Re^{m(m+1)/2}$ and $\cS^m$,
the dimension of $\Null(\nabla^2\phi(x))$ is $(m-k)(m-k+1)/2$, and hence the rank of $\nabla^2\phi(x)$ is
$m(m+1)/2 - (m-k)(m-k+1)/2$. To complete the proof, we next show that \eqref{null-space} holds
 by considering two cases $p\le -1$ or $-1<p<0$.

We start with the first case $p\le -1$. Notice that all entries of $S^{g,d}$ are nonnegative and
thus $M_1 \succeq 0$. Also, $M_2 \succeq 0$. It then follows from \eqref{KHess} and \eqref{Q} that
$h^T\nabla^2\phi(x)h=0$ if and only if
\beq \label{HMH}
\vec(H)^TM_1\vec(H)=0, \ \ \ \vec(H)^TM_2\vec(H)=0.
\eeq
By \eqref{Hess3}, the second equality of \eqref{HMH} becomes
   \begin{align}
      \tr(HX^{-1}K(K^TX^{-1}K)^{-p-1}K^TX^{-1}HX^{-1})& = 0, \nn
  \end{align}
  which is equivalent to
\beq \label{phesseq2}
\ba{c}
\tr(X^{-\frac12}HX^{-1}K(K^TX^{-1}K)^{-\frac{p+1}{2}}(K^TX^{-1}K)^{-\frac{p+1}{2}}K^TX^{-1}HX^{-\frac12}) = 0 \\ [4pt]
 \ \Leftrightarrow \ (K^TX^{-1}K)^{-\frac{p+1}{2}}K^TX^{-1}HX^{-\frac12} = 0 \ \Leftrightarrow \ K^TX^{-1}H = 0.
\ea
\eeq
Moreover, $K^TX^{-1}H=0$ implies that the first equality of \eqref{HMH} holds. Therefore, \eqref{HMH} holds if and only
if $K^TX^{-1}H=0$. It follows that \eqref{null-space} holds for $p \le -1$.

We next show that \eqref{null-space} also holds for $-1<p<0$. Indeed, for such $p$, all entries of $S^{g,d}$ are
negative and hence $-M_1 \succeq 0$.  Using Proposition~\ref{grad-Hess}, we see that $h^T\nabla^2\phi(x)h=0$ if and only if
\beq\label{HM1M2H}
\vec(H)^T(M_1+M_2)\vec(H)=0.
\eeq
We claim that
\beq \label{M1M2-ineq}
\frac12 \vec(H)^T M_2 \vec(H) \ge -\vec(H)^T M_1\vec(H).
\eeq
Indeed, letting $W = (K^TX^{-1}K)^{-1}$ and using Lemma~\ref{Schur2}, we have
  \begin{align*}
    W^{-1} = K^TX^{-1}K
    \ \Rightarrow \ \begin{pmatrix}
      X & K\\K^T &W^{-1}
    \end{pmatrix}\succeq 0
    \ \Rightarrow \ X\succeq KWK^T.
  \end{align*}
The latter relation together with the definitions of $M_2$, $G$ and \eqref{Hess3}
implies that
\beq \label{M2-ineq1}
\ba{l}
\frac12 \vec(H)^T M_2 \vec(H) \ = \ \tr(HX^{-1}HX^{-1}KW^{p+1}K^TX^{-1}) \\ [4pt]
\ = \ \tr([X^{-1}H(X^{-1}KW^{p+1}K^TX^{-1})^{\frac12}]^T X [X^{-1}H(X^{-1}KW^{p+1}K^TX^{-1})^{\frac12}]) \\ [4pt]
\ \ge \ \tr([X^{-1}H(X^{-1}KW^{p+1}K^TX^{-1})^{\frac12}]^T KWK^T [X^{-1}H(X^{-1}KW^{p+1}K^TX^{-1})^{\frac12}]) \\ [4pt]
\ = \ \tr(HX^{-1}KWK^TX^{-1}HX^{-1}KW^{p+1}K^TX^{-1})
\ea
\eeq
Let $Z=V^TK^TX^{-1}HX^{-1}KV$. Notice that $W = V\D(d^{-1})V^T$. Using this relation, the definition of $Z$ and
\eqref{vecformula}, we have
\[
\ba{l}
\tr(HX^{-1}KWK^TX^{-1}HX^{-1}KW^{p+1}K^TX^{-1}) \ = \ \tr(HX^{-1}KV\D(d^{-1})Z\D(d^{-p-1})V^TK^TX^{-1}) \\ [4pt]
\ = \ \tr(\D(d^{-1})Z\D(d^{-p-1})Z)  \ = \ \vec(Z)^T [\D(d^{-1})\otimes \D(d^{-p-1})]\vec(Z),
\ea
\]
which together with \eqref{M2-ineq1} yields
\beq \label{left-ineq}
\frac12 \vec(H)^T M_2 \vec(H) \ \ge \ \vec(Z)^T [\D(d^{-1})\otimes \D(d^{-p-1})]\vec(Z).
\eeq
Also, by the definitions of $M_1$, $Z$ and \eqref{vecformula}, we obtain that
\beq \label{right-ineq}
-\vec(H)^T M_1 \vec(H) = \vec(Z)^T \D({\bf vec}(-S^{g,d})) \vec(Z).
\eeq
 Since $1>p+1>0$ and $d_i > 0$ for all $i$, it is not hard to show that
  \[d_i^{-1}d_j^{-p-1} \ge -\frac{d_i^{-p-1}-d_j^{-p-1}}{d_i-d_j},\]
  whenever $d_i\neq d_j$. Thus, $\D(d^{-1})\otimes \D(d^{-p-1})\succeq \D({\bf vec}(-S^{g,d}))$, which
  together with \eqref{left-ineq} and \eqref{right-ineq} implies that \eqref{M1M2-ineq} holds. It then follows
from \eqref{M1M2-ineq}, \eqref{HM1M2H} and the fact $M_2 \succeq 0$ that $\vec(H)^T M_2 \vec(H)=0$. The rest of
proof is similar to the case $p \le -1$.
\end{proof}

\subsection{IP method for A-criterion} \label{sec:IP-A}

Recall from Section \ref{intro} that in $\cS^m_{++}$, the A-criterion $\Phi$ becomes
\begin{equation}\label{P2}
  \Phi(X) = \tr(K^TX^{-1}K)
\end{equation}
for some $K \in \Re^{m \times k}$ with full column rank. Since A-criterion is a special
case of $p$th mean criterion, the IP method discussed in Sections \ref{sec:IP} and
\ref{sec:p} can be suitably applied to solve problem \eqref{design} with
A-criterion. We next show that by exploiting the special structure, we can obtain a more
compact representation of the associated Hessian matrix that is used to compute Newton
direction for our IP method.



\begin{proposition}\label{A_grad-Hess}
Let $\Phi$ be defined in \eqref{P2} and the associated $\phi$ be defined in \eqref{phi}.
Then the gradient and Hessian of $\phi$ at any $x \in \svec(\cS^m_{++})$ are given by
 \begin{align}
    \nabla\phi(x)&= -\svec(X^{-1}KK^TX^{-1}),\label{AKgrad}\\
    \nabla^2\phi(x)&= 2X^{-1}\sotimes (X^{-1}KK^TX^{-1}),\label{AKHess}
  \end{align}
  where $X=\smat(x)$.
\end{proposition}

\begin{proof}
\eqref{AKgrad} follows immediately from \eqref{Kgrad} with $p=-1$. We
now prove \eqref{AKHess}. Let $x \in \svec(\cS^m_{++})$ be arbitrarily chosen, and let $X=\smat(x)$.
For all sufficiently small $h\in \Re^{m(m+1)/2}$, we observe $X+H \succ 0$,
where $H=\smat(h)$. In view of the definitions of $X$ and $H$, it then follows
from \eqref{AKgrad}, \eqref{X-inv} and \eqref{sotimes} that
\[
\ba{lcl}
\nabla\phi(x+h) - \nabla\phi(x) &=& -\svec\left((X+H)^{-1}KK^T(X+H)^{-1}-X^{-1}KK^TX^{-1}\right) \\ [4pt]
&=& \svec(X^{-1}HX^{-1}KK^TX^{-1} + X^{-1}KK^TX^{-1}HX^{-1}) + o(\svec(H)) \\ [4pt]
&=& 2X^{-1}\sotimes (X^{-1}KK^TX^{-1}) h+ o(h), \\
\ea
\]
which proves \eqref{AKHess}.
\end{proof}

\vgap

Since the A-criterion is a special case of the $p$th mean criterion, it follows from Proposition~\ref{rank_p} that
the rank of $\nabla^2 \phi(x)$ is also $m(m+1)/2 - (m-k)(m-k+1)/2$ for every $x \in \svec(\cS^m_{++})$.

\subsection{IP method for D-criterion}\label{sec:IP-D}

Recall from Section~\ref{intro} that in $\cS^m_{++}$, the
D-criterion $\Phi$ becomes
\begin{equation}\label{P3}
  \Phi(X) = \log\det(K^TX^{-1}K)
\end{equation}
for some $K\in \Re^{m\times k}$ with full column rank. It is easy to verify that
Assumption~\ref{assump} is satisfied. Hence, problem \eqref{design} with this
criterion can be suitably solved by the IP method studied in Section
\ref{sec:IP}. In the next proposition, we provide formulas for computing gradient and
Hessian of the associated function $\phi$ that are used in the IP method.
The proof is similar to that of Proposition~\ref{A_grad-Hess} and is thus omitted.

\begin{proposition}\label{D_grad-Hess}
Let $\Phi$ be defined in \eqref{P3} and the associated $\phi$ be defined in \eqref{phi}.
Then the gradient and Hessian of $\phi$ at any $x \in \svec(\cS^m_{++})$ are given by
   \begin{align}
    \nabla\phi(x)&= -\svec(X^{-1}KWK^TX^{-1}),\nonumber\\
    \nabla^2\phi(x)&= 2X^{-1}\sotimes (X^{-1}KW K^TX^{-1}) - (X^{-1}KWK^TX^{-1})\sotimes (X^{-1}KWK^TX^{-1}),\label{DKHess}
  \end{align}
  where $X=\smat(x)$ and $W = (K^TX^{-1}K)^{-1}$.
\end{proposition}

We next determine the rank of $\nabla^2\phi(X)$ at any $x \in \svec(\cS^m_{++})$.

\begin{proposition}\label{rank_D}
Let $\Phi$ be defined in \eqref{P3} and the associated $\phi$ be defined in \eqref{phi}.
Then the rank of $\nabla^2 \phi(x)$ is $m(m+1)/2 - (m-k)(m-k+1)/2$ for any $x\in \svec(\cS^m_{++})$.
\end{proposition}

\begin{proof}
Let $x\in \svec(\cS^m_{++})$ be arbitrarily chosen. Define $X=\smat(x)$. As in the proof of
Proposition~\ref{rank_p},  to determine the rank of $\nabla^2 \phi(x)$, it suffices to
know the dimension of $\Null(\nabla^2\phi(x))$. Notice that $\phi$ is a twice
differentiable convex function in $\svec(\cS^m_{++})$. Thus, $\nabla^2 \phi(x) \succeq 0$. It
implies that $h \in \Null(\nabla^2\phi(x))$ if and only if $h^T\nabla^2\phi(x)h=0$. In view of
\eqref{sotimes} and \eqref{DKHess}, it is not hard to verify that $h^T\nabla^2\phi(x)h=0$ if
and only if
\[
2\, \tr(HX^{-1}HX^{-1}KWK^TX^{-1}) - \tr(HX^{-1}KWK^TX^{-1}HX^{-1}KWK^TX^{-1}) = 0,
\]
where $H=\smat(h)$. In addition, we can observe that \eqref{M2-ineq1} also holds for $p=0$, and hence
\[
\tr(HX^{-1}HX^{-1}KWK^TX^{-1}) \ \ge \ \tr(HX^{-1}KWK^TX^{-1}HX^{-1}KWK^TX^{-1}).
\]
Furthermore,
\begin{align*}
& \ \tr(HX^{-1}KWK^TX^{-1}HX^{-1}KWK^TX^{-1}) \\
 = \ & \ \tr\left([(X^{-1}KWK^TX^{-1})^{\frac12}H]X^{-1}KWK^TX^{-1}[H(X^{-1}KWK^TX^{-1})^{\frac12}]\right) \ \ge \ 0.
\end{align*}
The above relations imply that $h^T\nabla^2\phi(x)h=0$ if and only if
\[
\tr(HX^{-1}HX^{-1}KWK^TX^{-1})=0,
\]
which together with definition of $W$ and the same arguments used in \eqref{phesseq2} implies that
\[
h^T\nabla^2\phi(x)h=0 \ \Leftrightarrow \ K^TX^{-1}H = 0.
\]
The rest of the proof follows similarly as that of Proposition~\ref{rank_p}.
\end{proof}

\section{Computational results}\label{sec:sim}

In this section, we conduct numerical experiments to test the performance of the IP method
discussed in this paper for solving problem~\eqref{design} with A-, D- and $p$th mean
criterion and also compare its performance with the multiplicative algorithm.

We develop Matlab codes for our IP method to solve
\eqref{design} with A-, D- and $p$th mean criterion. We also implement the
multiplicative algorithm in Matlab for solving \eqref{design} with
A-, D- and $p$th mean criterion. To benchmark the performance of
our IP method, we also report the computational results using a general SDP solver,
namely, SDPT3 \cite{TTT99,TTT03} (Version 4.0) on solving a linear SDP reformulation
of \eqref{design} with A-criterion (see \cite[Page~532]{FedLee00}) and a log-determinant SDP
reformulation of \eqref{design} with D-criterion (see \cite[Equation~(10)]{Papp10}). We shall
mention that it is not clear whether problem \eqref{design} with $p$th mean criterion
can be reformulated into a problem that can be efficiently solved by SDPT3. As SDPT3
implements an infeasible path-following algorithm, we project the approximate solution $w$ found
by SDPT3 onto the unit simplex to obtain an approximate optimal feasible solution for problem
\eqref{design} and the final objective value reported in our tests is based on the latter solution.
\footnote{Such projection makes a difference when SDPT3 terminates early at a solution that is highly infeasible,
which could be a consequence of ``near infeasibility" of the linear SDP reformulation;
see the first three rows of Table~\ref{table1}.}
All computations in this section are performed in Matlab 7.14.0 (2012a) on a workstation with
an Intel Xeon E5410 CPU (2.33 GHz) and 8GB RAM running Red Hat Enterprise Linux (kernel 2.6.18).

For our IP method, we set
$\tw^0=\frac{1}{n}e\in \Re^{n-1}$, $\mu_1=10$, $\beta=\gamma=0.5$,
$\sigma=0.1$ and $\eta=0.95$. In addition, we set
$\epsilon(\mu) = \max\{\mu,1e-10\}$
and terminate the algorithm once $\mu_k\le 1e-10$.
On the other hand, for
the multiplicative algorithm, similarly as in \cite{Yu10a}, we set $\lambda=1$, $w^0=\frac{1}{n}e\in \Re^n$, and
terminate the algorithm when it reaches $10000$ iterations or
\begin{equation*}
  \max_{1\le i\le n} d_i(w^k) \le (1+\delta) \sum_{i=1}^nw^k_id_i(w^k)
\end{equation*}
holds with $\delta = 2e-4$, where $d_i(w)$ is defined in \eqref{power}
\footnote{We also tried $\delta = 1e-4$, but the multiplicative algorithm tends to take a long time for relatively little improvement on some instances.}.
Furthermore,
for SDPT3, we use the default tolerance.
Finally, we use the mex files skron, smat and svec from the SDPT3 package for
efficient operations on symmetric matrices in our implementation of the
IP method and the multiplicative algorithm.

In our tests below, we consider the following four design spaces:
\begin{align*}
\begin{array}{rl}
  \chi_1(n)&=\{x_i=(e^{-s_i},s_ie^{-s_i},e^{-2s_i},s_ie^{-2s_i})^T, 1\le i\le
  n\},\\
  \chi_2(n)&=\{x_i=(1,s_i,s_i^2,s_i^3)^T, 1\le i\le
  n\},\\
  \chi_3(n)&=\{x_{(i-1)\lceil\sqrt{n}\rceil+j}=(1,r_i,r_i^2,t_j,r_it_j)^T, 1\le i,j\le \lceil\sqrt{n}\rceil\},\\
  \chi_4(n)&=\{x_i=(t_i,t_i^2,\sin(2 \pi t_i),\cos(2\pi t_i))^T, 1\le i
  \le n\},
  \end{array}
\end{align*}
where $s_i=\frac{3i}{n}$, $r_i=\frac{2i}{n}-1$ and $t_i=\frac{i}{n}$.
The space $\chi_1(n)$ represents the linearization of a
compartmental model \cite{ACHJ93}.
The space $\chi_2(n)$ corresponds to polynomial regression.
The third space, as described in \cite{Yu10b}, represents a response surface
with a nonlinear effect and an interaction, while the fourth space
is the quadratic/trigonometric example proposed in \cite{Wyn72}. The test sets $\chi_1$,
$\chi_3$ and a variant of the test set $\chi_2$ are also used in \cite{Yu10b}.

In our first test, for each design space, we set $A_i =x_ix_i^T$ for $i=1,\ldots,n$,
with $n = 10000$, $50000$, $100000$ for $\chi_1$, $\chi_2$, $\chi_4$,
and $n = 10000$, $40000$, $90000$ for $\chi_3$. For each $n$ and each design space, we randomly
generate $30$ different matrices $K\in \Re^{m\times 3}$ (i.e., we set $k = 3$), each having
i.i.d. Gaussian entries of mean $0$ and variance $1$.
We then apply our IP method and the multiplicative algorithm to solve
problem~\eqref{design} with A-, D- and $p$th mean criterion on these instances
and also apply SDPT3 to solve \eqref{design} with A- and D-criterion. The
computational results averaged over the $30$ instances are reported in
Tables~\ref{table1}--\ref{table3}. In particular, the performance of our
IP method, the multiplicative algorithm and SDPT3 are reported under
the columns named ``IP'', ``MUL'' and ``SDPT3'', respectively. In addition,
the CPU time abbreviated as ``cpu'' is in seconds and the objective value
abbreviated as ``obj'' is rounded off to six significant digits. We see that our
IP method significantly outperforms the multiplicative algorithm in terms of CPU time,
and gives a smaller objective value in all instances. Moreover, our IP method also
outperforms SDPT3 in CPU time and gives a smaller objective value in most instances.
Furthermore, it is worth pointing out that SDPT3 reports infeasibility and hence {\em early terminates}
when solving some instances for $\chi_1$ with A-criterion, possibly due to
bad scaling of $\cM(w)$. This accounts for its
significantly larger objective values in Table~\ref{table1} corresponding to $\chi_1$.
Finally, for $p$th mean criterion with $p < -1$, our IP method achieves
significantly better objective values than the multiplicative algorithm, where
the objective value of the latter algorithm is chosen to be the minimum over all
iterations (see Table~\ref{table3}). This phenomenon is
actually not surprising since the multiplicative algorithm is only
known to converge for $p \in (-1,0)$, but it may not converge when
$p < -1$.

\begin{table}[t!]
\begin{center}
{\small \caption{\small Computational results for A-criterion with random $K$}\label{table1}
\centering
\begin{tabular}{|cc||ccc||ccc|}
\hline
\multicolumn{2}{|c||}{} & \multicolumn{3}{c||}{cpu}  &
\multicolumn{3}{c|}{obj}
\\
$\chi_{_i}$& $n$ & \multicolumn{1}{c}{MUL}  &
\multicolumn{1}{c}{IP}& \multicolumn{1}{c||}{SDPT3}  &
\multicolumn{1}{c}{MUL}  &
\multicolumn{1}{c}{IP}& \multicolumn{1}{c|}{SDPT3}
\\ \hline
  1 & 10000  &  13.76   &   0.69  &   1.93  &         193041  &         191410  &         211735  \\
  1 & 50000  &  62.13   &   3.75  &  10.64  &         154584  &         153219  &         172514  \\
  1 & 100000 &  135.26  &   7.38  &  19.22  &         208599  &         206787  &         242633  \\
  2 & 10000  &  17.69   &   0.77  &   1.90  &        215.754  &        212.356  &        212.356  \\
  2 & 50000  &  89.89   &   4.22  &  10.94  &        188.509  &         185.55  &        185.551  \\
  2 & 100000 &  163.28  &   8.12  &  21.83  &        242.414  &        237.823  &        237.824  \\
  3 & 10000  &  33.74   &   1.02  &   2.36  &        54.8551  &        54.7332  &        54.7332  \\
  3 & 40000  &  140.85  &   5.24  &  14.17  &        49.0008  &        48.9784  &        48.9791  \\
  3 & 90000  &  322.52  &  10.75  &  35.50  &        50.8124  &        50.7906  &        50.7906  \\
  4 & 10000  &  13.42   &   0.90  &   1.90  &        572.779  &        558.088  &        558.088  \\
  4 & 50000  &  58.41   &   4.46  &   9.89  &        501.924  &         487.99  &        487.991  \\
  4 & 100000 &  139.62  &   9.03  &  20.36  &        343.827  &        337.003  &        337.023  \\ \hline
\end{tabular}}
\end{center}
\end{table}

\begin{table}[t!]
\begin{center}
{\small \caption{\small Computational results for D-criterion with random $K$}\label{table11}
\centering
\begin{tabular}{|cc||ccc||ccc|}
\hline
\multicolumn{2}{|c||}{} & \multicolumn{3}{c||}{cpu}  &
\multicolumn{3}{c|}{obj}
\\
$\chi_{_i}$& $n$ & \multicolumn{1}{c}{MUL}  &
\multicolumn{1}{c}{IP}& \multicolumn{1}{c||}{SDPT3}  &
\multicolumn{1}{c}{MUL}  &
\multicolumn{1}{c}{IP}& \multicolumn{1}{c|}{SDPT3}
\\ \hline
  1 & 10000  &   1.47  &   0.95  &   1.47  &        19.7352  &        19.7347  &        19.7356  \\
  1 & 50000  &   5.40  &   4.73  &   6.03  &        19.9312  &        19.9307  &         19.933  \\
  1 & 100000 &  14.17  &   9.31  &  12.34  &        19.7973  &        19.7968  &        19.7987  \\
  2 & 10000  &   2.10  &   0.79  &   1.57  &        5.95269  &        5.95229  &         5.9523  \\
  2 & 50000  &  20.60  &   4.07  &   6.78  &        5.30436  &         5.3039  &         5.3039  \\
  2 & 100000 &  51.43  &   8.33  &  13.32  &        5.08652  &        5.08608  &        5.08609  \\
  3 & 10000  &   4.31  &   1.05  &   1.77  &        6.58713  &        6.58694  &        6.58694  \\
  3 & 40000  &  18.86  &   4.28  &   9.11  &        6.65124  &        6.65104  &        6.65103  \\
  3 & 90000  &  66.40  &   9.93  &  21.48  &        6.74346  &        6.74327  &        6.74382  \\
  4 & 10000  &   1.67  &   0.90  &   1.41  &        7.40587  &        7.40535  &        7.40535  \\
  4 & 50000  &  13.46  &   4.35  &   6.09  &        7.65401  &         7.6535  &         7.6535  \\
  4 & 100000 &  39.01  &   7.90  &  12.03  &        8.66619  &        8.66575  &        8.66574  \\ \hline
\end{tabular}}
\end{center}
\end{table}

\begin{table}[t!]
\begin{center}
{\small \caption{\small Computational results for $p$th mean criterion with random $K$ for some $p\in (-1,0)$}\label{table2}
\centering
\begin{tabular}{|cc||cc||cc||cc||cc|}
\hline
\multicolumn{2}{|c||}{} & \multicolumn{4}{c||}{$p = -0.25$}  &
\multicolumn{4}{c|}{$p = -0.75$}\\
\multicolumn{2}{|c||}{} & \multicolumn{2}{c}{cpu}  &
\multicolumn{2}{c||}{obj}& \multicolumn{2}{c}{cpu}  &
\multicolumn{2}{c|}{obj}
\\
$\chi_{_i}$& $n$  &\multicolumn{1}{c}{MUL}  &
\multicolumn{1}{c}{IP}&
\multicolumn{1}{c}{MUL}  &\multicolumn{1}{c||}{IP}
&\multicolumn{1}{c}{MUL}  &
\multicolumn{1}{c}{IP}&\multicolumn{1}{c}{MUL}  &\multicolumn{1}{c|}{IP}
\\ \hline
  1 & 10000  &   6.39  &   0.85  &   25.4567  &  25.4558  &   9.17  &   0.79  &   5187.73  &  5187.23      \\
  1 & 50000  &  36.84  &   4.18  &   25.1902  &  25.1894  &  45.72  &   4.09  &   7128.51  &  7126.32      \\
  1 & 100000 &  82.59  &   8.29  &   25.1312  &  25.1304  &  118.04 &   8.11  &   7207.59  &  7205.68      \\
  2 & 10000  &   6.19  &   0.76  &   5.68067  &  5.68046  &  13.67  &   0.79  &   46.4144  &  46.4008      \\
  2 & 50000  &  28.10  &   3.89  &   6.00911  &  6.00886  &  73.65  &   4.28  &   58.2028  &  58.1903      \\
  2 & 100000 &  71.37  &   8.00  &   6.12458  &  6.12434  &  152.84 &   8.30  &   60.9256  &  60.9108      \\
  3 & 10000  &   5.47  &   0.98  &   5.58387  &  5.58379  &   3.73  &   1.07  &   24.8691  &   24.868      \\
  3 & 40000  &  18.97  &   4.18  &   5.56727  &  5.56718  &  17.21  &   4.66  &   23.5463  &  23.5451      \\
  3 & 90000  &  58.35  &   9.91  &   5.45907  &  5.45899  &  56.66  &  10.98  &   24.7679  &  24.7664      \\
  4 & 10000  &   1.84  &   0.90  &   7.30484  &  7.30456  &   2.88  &   0.90  &   118.871  &  118.859      \\
  4 & 50000  &   8.84  &   4.63  &   7.27622  &  7.27589  &  10.98  &   4.97  &   108.079  &  108.066      \\
  4 & 100000 &  31.14  &   8.81  &   7.30129  &  7.30102  &  45.27  &   9.70  &   128.676  &  128.662      \\
\hline
\end{tabular}}
\end{center}
\end{table}

\begin{table}[t!]
\begin{center}
{\small \caption{\small Computational results for $p$th mean criterion with random $K$ for some $p < -1$}\label{table3}
\hspace*{-1cm}
\centering
\begin{tabular}{|cc||cc||cc||cc||cc|}
\hline
\multicolumn{2}{|c||}{} & \multicolumn{4}{c||}{$p = -1.1$}  &
\multicolumn{4}{c|}{$p = -1.2$}\\
\multicolumn{2}{|c||}{} & \multicolumn{2}{c}{cpu}  &
\multicolumn{2}{c||}{obj}& \multicolumn{2}{c}{cpu}  &
\multicolumn{2}{c|}{obj}
\\
$\chi_{_i}$& $n$ &\multicolumn{1}{c}{mul}  &
\multicolumn{1}{c}{IP}&
\multicolumn{1}{c}{mul}  &\multicolumn{1}{c||}{IP}
&\multicolumn{1}{c}{mul}  &
\multicolumn{1}{c}{IP}&
\multicolumn{1}{c}{mul}  &\multicolumn{1}{c|}{IP}
\\ \hline
  1 & 10000  &   5.07  &   0.70  &    611960  &    602294  &   4.66  &   0.68  &    1.46813e+06  &    1.43891e+06       \\
  1 & 50000  &  20.64  &   3.59  &    541355  &    532904  &  19.71  &   3.59  &     2.0649e+06  &    2.02777e+06       \\
  1 & 100000 &  48.85  &   7.47  &    371942  &    365201  &  51.33  &   7.35  &    1.79042e+06  &    1.75803e+06       \\
  2 & 10000  &   6.25  &   0.80  &   373.376  &   359.802  &   5.34  &   0.79  &        650.345  &        629.288       \\
  2 & 50000  &  21.38  &   4.12  &   492.463  &   476.123  &  21.15  &   4.13  &        667.047  &            645       \\
  2 & 100000 &  61.53  &   8.44  &   302.083  &    288.88  &  62.32  &   8.41  &        539.641  &        514.087       \\
  3 & 10000  &  19.16  &   1.11  &   74.7421  &   71.4397  &  20.40  &   1.18  &         95.224  &         88.142       \\
  3 & 40000  &  71.26  &   4.76  &   69.2354  &   65.8478  &  68.95  &   4.79  &        127.857  &        116.933       \\
  3 & 90000  &  204.48 &  11.78  &   69.2994  &   65.8143  &  160.07 &  11.85  &        109.287  &        100.475       \\
  4 & 10000  &   6.75  &   0.92  &    961.75  &   910.571  &   7.17  &   0.96  &        1640.19  &        1524.98       \\
  4 & 50000  &  27.46  &   4.91  &   903.773  &   846.954  &  36.12  &   5.02  &         1631.8  &        1520.71       \\
  4 & 100000 &  75.39  &   9.85  &   824.269  &   776.036  &  75.90  &   9.82  &        1710.28  &         1596.1       \\
\hline
\end{tabular}}
\end{center}
\end{table}

In our second test, we consider the case when $K = I$.
The instances used in this test are the same as those in the first test except $K=I$.
We also apply our IP method and the multiplicative algorithm to solve problem~\eqref{design}
 with A-, D- and $p$th mean criterion on these instances and apply SDPT3 to solve \eqref{design}
with A- and D-criterion. The computational results are reported in Tables~\ref{table4}--\ref{table6}.
We again observe that our IP method outperforms the multiplicative algorithm in terms of objective
value in all instances, and is generally much faster on large instances.
Furthermore, our IP method is usually faster than SDPT3 and produces comparable
or smaller objective values.

\begin{table}[t!]
\begin{center}
{\small \caption{\small Computational results for A-criterion with $K = I$}\label{table4}
\centering
\begin{tabular}{|cc||ccc||ccc|}
\hline
\multicolumn{2}{|c||}{} & \multicolumn{3}{c||}{cpu}  &
\multicolumn{3}{c|}{obj}
\\
$\chi_{_i}$& $n$ & \multicolumn{1}{c}{mul}  &
\multicolumn{1}{c}{IP}& \multicolumn{1}{c||}{SDPT3}  &
\multicolumn{1}{c}{mul}  &
\multicolumn{1}{c}{IP}& \multicolumn{1}{c|}{SDPT3}
\\ \hline
  1 & 10000  &  13.69  &   0.74  &   2.29  &        54286.3  &        53848.3  &        53848.4 \\
  1 & 50000  &  62.08  &   4.17  &  12.23  &        54245.2  &        53807.3  &        54103.8 \\
  1 & 100000 &  133.65 &   7.37  &  27.46  &        54240.1  &        53802.1  &        54103.8 \\
  2 & 10000  &  16.53  &   0.81  &   1.82  &        73.4521  &        72.4443  &        72.4443 \\
  2 & 50000  &  75.99  &   4.26  &  11.03  &         73.391  &         72.385  &        72.3853 \\
  2 & 100000 &  164.18 &   8.60  &  20.23  &        73.3837  &        72.3778  &        72.3777 \\
  3 & 10000  &   1.58  &   0.93  &   2.13  &        21.6203  &        21.6191  &        21.6191 \\
  3 & 40000  &  12.81  &   4.38  &  11.38  &        21.2826  &        21.2812  &        21.2812 \\
  3 & 90000  &  36.66  &   9.14  &  30.21  &        21.1721  &        21.1706  &        21.1706 \\
  4 & 10000  &  12.84  &   0.96  &   1.58  &        174.279  &        170.775  &        170.775 \\
  4 & 50000  &  59.76  &   5.19  &   9.51  &        174.276  &        170.775  &        170.775 \\
  4 & 100000 &  128.73 &   9.93  &  17.13  &        174.277  &        170.775  &        170.776 \\
\hline
\end{tabular}}
\end{center}
\end{table}

\begin{table}[t!]
\begin{center}
{\small \caption{\small Computational results for D-criterion with $K = I$}\label{table41}
\centering
\begin{tabular}{|cc||ccc||ccc|}
\hline
\multicolumn{2}{|c||}{} & \multicolumn{3}{c||}{cpu}  &
\multicolumn{3}{c|}{obj}
\\
$\chi_{_i}$& $n$ & \multicolumn{1}{c}{mul}  &
\multicolumn{1}{c}{IP}& \multicolumn{1}{c||}{SDPT3}  &
\multicolumn{1}{c}{mul}  &
\multicolumn{1}{c}{IP}& \multicolumn{1}{c|}{SDPT3}
\\ \hline
  1 & 10000  &   1.11  &   1.02  &   0.87  &        20.5125  &        20.5119  &        20.5125  \\
  1 & 50000  &   4.86  &   4.67  &   3.86  &        20.5098  &        20.5091  &        20.5091  \\
  1 & 100000 &  14.77  &   9.13  &   7.59  &        20.5094  &        20.5087  &        20.5088  \\
  2 & 10000  &   1.92  &   0.74  &   1.01  &       0.410745  &       0.410221  &        0.41022  \\
  2 & 50000  &  16.74  &   3.80  &   4.75  &       0.409964  &       0.409267  &        0.40926  \\
  2 & 100000 &  55.28  &   6.96  &   8.89  &       0.409795  &       0.409154  &       0.409145  \\
  3 & 10000  &   1.76  &   0.89  &   1.16  &        5.14292  &        5.14267  &        5.14267  \\
  3 & 40000  &  15.90  &   3.99  &   6.53  &        5.08236  &        5.08212  &        5.08211  \\
  3 & 90000  &  47.07  &   8.70  &  15.38  &        5.06226  &        5.06202  &        5.06201  \\
  4 & 10000  &   1.35  &   1.02  &   0.94  &        7.25257  &        7.25189  &        7.25189  \\
  4 & 50000  &  11.16  &   5.04  &   4.17  &        7.25253  &         7.2519  &        7.25189  \\
  4 & 100000 &  35.09  &   9.86  &   8.14  &        7.25246  &         7.2519  &        7.25189  \\
\hline
\end{tabular}}
\end{center}
\end{table}

\begin{table}[t!]
\begin{center}
{\small \caption{\small Computational results for $p$th mean criterion with $K = I$ for some $p\in (-1,0)$}\label{table5}
\centering
\begin{tabular}{|cc||cc||cc||cc||cc|}
\hline
\multicolumn{2}{|c||}{} & \multicolumn{4}{c||}{$p = -0.25$}  &
\multicolumn{4}{c|}{$p = -0.75$}\\
\multicolumn{2}{|c||}{} & \multicolumn{2}{c}{cpu}  &
\multicolumn{2}{c||}{obj}& \multicolumn{2}{c}{cpu}  &
\multicolumn{2}{c|}{obj}
\\
$\chi_{_i}$& $n$  &\multicolumn{1}{c}{mul}  &
\multicolumn{1}{c}{IP}&
\multicolumn{1}{c}{mul}  &\multicolumn{1}{c||}{IP}
&\multicolumn{1}{c}{mul}  &
\multicolumn{1}{c}{IP}&\multicolumn{1}{c}{mul}  &\multicolumn{1}{c|}{IP}
\\ \hline
  1 & 10000  &  7.48  &   0.91  &    23.3728  &    23.372   &  3.53  &   0.82  &   3635.71  &   3635.29         \\
  1 & 50000  & 42.29  &   4.21  &    23.3683  &   23.3675   & 24.12  &   4.37  &   3633.58  &    3633.2         \\
  1 & 100000 & 91.80  &   8.29  &    23.3677  &    23.367   & 57.01  &   8.86  &   3633.31  &   3632.94         \\
  2 & 10000  &  3.43  &   0.74  &    5.58855  &   5.58838   &  2.55  &   0.80  &   27.4836  &   27.4811         \\
  2 & 50000  & 20.56  &   4.00  &    5.58796  &   5.58771   & 11.69  &   4.14  &   27.4691  &   27.4653         \\
  2 & 100000 & 69.43  &   7.67  &    5.58785  &   5.58763   & 37.49  &   8.60  &    27.467  &   27.4634         \\
  3 & 10000  &   1.65 &   0.77  &    6.70457  &   6.70448   &   1.56 &   0.99  &   14.1435  &   14.1429         \\
  3 & 40000  &  14.46 &   4.24  &    6.68235  &   6.68225   &  13.22 &   3.82  &   13.9841  &   13.9834         \\
  3 & 90000  &  42.01 &   8.24  &      6.675  &   6.67491   &  37.85 &   9.25  &   13.9318  &   13.9311         \\
  4 & 10000  &  1.75  &   0.92  &    7.25984  &   7.25955   &  1.43  &   0.92  &   52.2922  &    52.286         \\
  4 & 50000  &  8.97  &   4.58  &    7.25988  &   7.25956   &  6.05  &   4.52  &   52.2937  &    52.286         \\
  4 & 100000 & 30.50  &   8.85  &    7.25983  &   7.25957   & 20.67  &   9.20  &   52.2927  &   52.2861         \\
\hline
\end{tabular}}
\end{center}
\end{table}

\begin{table}[t!]
\begin{center}
{\small \caption{\small Computational results for $p$th mean criterion with $K = I$ for some $p < -1$}\label{table6}
\centering
\begin{tabular}{|cc||cc||cc||cc||cc|}
\hline
\multicolumn{2}{|c||}{} & \multicolumn{4}{c||}{$p = -1.1$}  &
\multicolumn{4}{c|}{$p = -1.2$}\\
\multicolumn{2}{|c||}{} & \multicolumn{2}{c}{cpu}  &
\multicolumn{2}{c||}{obj}& \multicolumn{2}{c}{cpu}  &
\multicolumn{2}{c|}{obj}
\\
$\chi_{_i}$& $n$  &\multicolumn{1}{c}{mul}  &
\multicolumn{1}{c}{IP}&
\multicolumn{1}{c}{mul}  &\multicolumn{1}{c||}{IP}
&\multicolumn{1}{c}{mul}  &
\multicolumn{1}{c}{IP}&\multicolumn{1}{c}{mul}  &\multicolumn{1}{c|}{IP}
\\ \hline
  1 & 10000   &  4.62  &   0.72  &  162818  &    159210   &  4.64  &   0.67  &     485415  &     471459    \\
  1 & 50000   & 18.15  &   3.64  &  162740  &    159077   & 18.28  &   3.63  &     482380  &     471030    \\
  1 & 100000  & 49.28  &   7.47  &  162732  &    159060   & 47.57  &   7.56  &     485149  &     470975    \\
  2 & 10000   &  4.59  &   0.79  & 108.922  &   108.171   &  4.65  &   0.80  &    165.133  &    162.297    \\
  2 & 50000   & 18.44  &   4.19  & 109.588  &   108.072   & 19.23  &   4.27  &    164.314  &    162.134    \\
  2 & 100000  & 47.95  &   8.66  & 109.495  &    108.06   & 45.78  &   8.87  &    165.458  &    162.114    \\
  3 & 10000   &  36.72 &   1.02  & 25.9565  &   25.7793   &  36.40 &   1.04  &    31.8264  &    30.8276    \\
  3 & 40000   &  142.62&   4.17  &  25.599  &   25.3307   &  139.85&   4.50  &    31.5254  &    30.2362    \\
  3 & 90000   &  328.00&   9.60  & 25.5115  &   25.1841   &  322.93&   9.68  &      31.46  &    30.0431    \\
  4 & 10000   &  6.51  &   0.95  & 297.604  &   277.597   &  7.99  &   0.89  &    497.138  &        453    \\
  4 & 50000   & 27.17  &   4.83  & 297.686  &   277.597   & 33.88  &   4.89  &    497.287  &        453    \\
  4 & 100000  & 63.40  &   9.57  & 297.696  &   277.597   & 81.00  &  10.02  &    497.306  &        453    \\
\hline
\end{tabular}}
\end{center}
\end{table}

\section{Concluding remarks}
\label{conclude}

In this paper we propose an IP method for
solving problem~\eqref{design} with a broad class of convex optimality
criteria and establish its global convergence. We demonstrate
how the Newton direction can be efficiently computed when the method
is applied to \eqref{design} with classical optimality criteria.
Our computational results show that the IP method outperforms the widely used
multiplicative algorithm in both speed and solution quality. The codes for this paper,
including our implementation of the multiplicative algorithm and our codes generating
inputs for SDPT3, are available online at www.math.sfu.ca/$\sim$zhaosong.

Finally, we would like to remark that the performance of our IP method depends on whether
the Newton direction can be computed accurately and efficiently. In our implementation,
we observe that for $p$th mean criterion with large $|p|$, as well as for
the design space $ \{x_i=(1,s_i,s_i^2,s_i^3,s_i^4)^T, 1\le i\le n\}$ with $n \ge 50000$
and some random $K\in \Re^{m\times 3}$, the Newton direction cannot be computed accurately
due to numerical errors and hence our IP method fails to terminate with a good approximate
solution, compared with the multiplicative algorithm. Indeed, it is known \cite{Wri95,Wri99} that
the performance of a barrier method deteriorates as $\mu\rightarrow 0$.
It is conceivable that such issues would not arise if a primal-dual IP method was used instead. However,
it is much more involved to develop a primal-dual IP method for solving \eqref{design}: since
the feasible set of \eqref{design} is not closed in general,
one would have to develop a primal-dual IP method on an equivalent nonlinear semidefinite programming reformulation of \eqref{design}.
We leave this as a future research direction.

\section*{Appendix}


We present the proof of Theorem~\ref{convergence} in this appendix.

\begin{proof}
In this proof, we denote by $\tilde w(\mu,v)$ the vector obtained from $w(\mu,v)$ by dropping the last
entry for all $(\mu, v)\in \Re_{++} \times \Re^{n-1}$. Notice that $\tilde w(\mu,v)$ is the unique optimal solution of
\eqref{KKT-p}.

We now prove part (a). Let
\beq \label{barf}
\barf^* := \inf_w\{\Phi_\cM(w):\; e^Tw = 1, w > 0 \}.
\eeq

We first show that $\lim\limits_{(\mu,v) \rightarrow (0_+,0)} \Phi_\cM( w(\mu,v)) = \barf^*$.

Given an arbitrary $\eps>0$, there exists a positive $\tw$ satisfying $e^T\tw < 1$ such that
$f(\tw) < \barf^* + \eps/2$.  
Then we have that for any $v \in \Re^{n-1}$,
\begin{align}\label{ineq}
   f_\mu(\tilde w(\mu,v))-v^T \tilde w(\mu,v)\le f_\mu(\tilde w)- v^T
   \tilde w.
\end{align}
On the other hand, note that $\tilde w(\mu,v) > 0$ and $e^T \tilde w(\mu,v) <1$.   Hence,
$$-\sum_{i=1}^{n-1}\log(\tw_i(\mu,v))-\log\left(1-e^T\tw(\mu,v)\right)>0$$
and $f(\tilde w(\mu,v))\ge \barf^*$. In view of these inequalities, \eqref{ineq} and
the fact that $\|\tilde w(\mu,v)\|_1 \le 1$ and $\|\tw\|_1 \le 1$, one can obtain that
for any $(\mu, v)\in \Re_{++} \times \Re^{n-1}$,
\begin{align*}
  \barf^*\le f(\tilde w(\mu,v))& =  f_\mu(\tilde w(\mu,v)) + \mu\sum_{i=1}^{n-1}\log(\tw_i(\mu,v)) + \mu\log\left(1-e^T\tw(\mu,v)\right) \\
  & \le f_\mu(\tilde w(\mu,v)) \ \le \ f_\mu(\tilde w) + v^T \tilde w(\mu,v) - v^T \tilde w \\
& \le f(\tilde w)-\mu\sum_{i=1}^{n-1}\log(\tilde w_i)-\mu \log\left(1-e^T\tilde w\right)+2\|v\|_\infty\\
&\le \barf^*+\frac{\eps}{2}-\mu\sum_{i=1}^{n-1}\log(\tilde w_i)-\mu
\log\left(1-e^T\tilde w\right)+2\|v\|_\infty.
\end{align*}
Thus, there exists some $\delta>0$ such that $\barf^*\le f(\tilde
w(\mu,v)) \le \barf^* + \eps$ whenever $\|(\mu,v)\|< \delta$, $\mu>0$.
Hence, $\Phi_\cM(w(\mu,v)) = f(\tilde w(\mu,v)) \to \barf^*$ as $(\mu,v) \rightarrow
(0_+,0)$.

We next show that $f^* = \barf^*$. Clearly, $f^* \le \barf^*$. We now suppose for contradiction
that $f^* < \barf^*$. By the definitions of $f^*$ and $\barf^*$, there exist $w^1$ and $w^2$ which are feasible points
of \eqref{design} and \eqref{barf}, respectively, so that $\Phi_\cM(w^1) < (f^*+\barf^*)/2$ and $\Phi_\cM(w^2)  <  \barf^* + (\barf^*-f^*)/2$.
Let $w = (w^1+w^2)/2$. Clearly, $w > 0$, $e^T w = 1$ and ${\rm Range}(K)\subseteq {\rm Range}(\cM(w))$ due to $\cM(w) \succ 0$. By convexity of $\Phi$ in
$\cS^m_{+}(K)$, we obtain that $\Phi_\cM(w) \le (\Phi_\cM(w^1) + \Phi_\cM(w^2))/2 < \barf^*$, which is a contradiction to the definition
of $\barf^*$. Thus, $\lim\limits_{(\mu,v) \rightarrow (0_+,0)} \Phi_\cM( w(\mu,v)) = \barf^* = f^*$.


Now suppose that $w^*$ is an accumulation point of $w(\mu,v)$ as $(\mu,v) \rightarrow (0_+,0)$. We next show that
$w^*$ is an optimal solution of \eqref{design}. Indeed, it follows from \eqref{phix}
that for any feasible point $w$ of \eqref{design},
\begin{equation}\label{infformula2}
  \Phi_\cM(w) = \inf_U\left\{\Psi(U):\; \cM(w)\succeq KUK^T, U\succ 0\right\}.
\end{equation}
In view of \eqref{infformula2}, for each $(\mu,v)\in \Re_{++}\times \Re^{n-1}$, there exists $U(\mu,v)\succ 0$ such that
  \begin{equation}\label{choice1}
  \Phi_\cM(w(\mu,v))+\|(\mu,v)\|>\Psi(U(\mu,v))\ \ \ {\rm and}\ \ \ \cM(w(\mu,v))\succeq KU(\mu,v)K^T.
  \end{equation}
  From the second relation in \eqref{choice1}, we see that $\tr(\cM(w(\mu,v)))\ge \lambda_{\min}(K^TK){\rm
  tr}(U(\mu,v))$, from which it follows that $U(\mu,v)$ is bounded and thus it has an accumulation point
  as $(\mu,v) \rightarrow (0_+,0)$. Let
  $U^*$ be such an accumulation point. In view of the first
  relation in \eqref{choice1} and the assumption on $\Psi$, we see that $U^*\succ 0$. Moreover, we obtain
  by taking limit in \eqref{choice1} upon $(\mu,v)\rightarrow (0_+,0)$ that
\beq \label{limit-pt}
\lim\limits_{(\mu,v)\rightarrow (0_+,0)}\Phi_\cM(w(\mu,v))\ge\Psi(U^*), \ \ \
\cM(w^*)\succeq KU^*K^T.
\eeq
The second relation in \eqref{limit-pt} together with Lemma~\ref{Schur2} implies that
   \begin{align*}
    &\ \cM(w^*)\succeq KU^*K^T \ \Rightarrow \ \begin{pmatrix}
      {U^*}^{-1}&K^T\\K&\cM(w^*)
    \end{pmatrix}\succeq 0 \
    \Rightarrow \ {\rm Range}(K)\subseteq {\rm Range}(\cM(w^*)).
  \end{align*}
  Hence, $w^*$ is a feasible point of \eqref{design}. In view of \eqref{infformula2}, the first relation in
\eqref{limit-pt} and the result $\lim\limits_{(\mu,v) \rightarrow (0_+,0)} \Phi_\cM( w(\mu,v)) = f^*$, we have
  \[
   \Phi_\cM(w^*) \le \Psi(U^*) \le  \lim_{(\mu,v)\rightarrow(0_+,0)}  \Phi_\cM( w(\mu,v)) = f^*.
  \]
Thus, $w^*$ is an optimal solution of \eqref{design}. This proves part (a).

We next show that part (b) holds. Let $w^\star$ be an optimal solution of
\eqref{design} with maximum cardinality. Then it follows
immediately from assumption that $\cM(w^\star)\succ 0$. Thus, there exists a
corresponding Lagrange multiplier $u^\star$ so that $(w^\star,u^\star)$
satisfies \eqref{KKToriginal}.  Let
$\tw^\star$ be the vector obtained from $w^\star$ by dropping the last entry. In view
of \eqref{Bh} and the first equation of \eqref{KKToriginal} and \eqref{KKTfmuapprox},
we observe that for any $(\mu,v) \in \Xi_C$,
\begin{align*}
  &(w(\mu,v)-w^\star)^T(u(\mu,v)-u^\star)\\
  =\ &(P\tilde w(\mu,v)-P\tilde w^\star)^T(u(\mu,v)-u^\star) \\
  =\ &(\tilde w(\mu,v)-\tilde w^\star)^TP^T(\nabla\Phi_\cM(w(\mu,v))-\nabla\Phi_\cM(w^\star))-(\tilde w(\mu,v)-\tilde w^\star)^T v\\
  =\ &(w(\mu,v)- w^\star)^T(\nabla\Phi_\cM(w(\mu,v))-\nabla\Phi_\cM(w^\star))-(\tilde w(\mu,v)-\tilde w^\star)^T v\\
  \ge\ & -2C\mu,
\end{align*}
where the last inequality holds since $\Phi$ is convex in
$\cS^m_{++}$, $w(\mu,v), w^\star\in \Omega$ and $\|v\|_\infty< C\mu$.
Using this inequality and the third equation in \eqref{KKToriginal}
and \eqref{KKTfmuapprox}, we see that
\begin{align}\label{bound0}
  {w^\star}^Tu(\mu,v)+{w(\mu,v)}^Tu^\star& \ \le \ {w^\star}^Tu^\star+{w(\mu,v)}^Tu(\mu,v)+2C\mu
  \ = \ (2C+n)\mu.
\end{align}
Dividing both sides of the above inequality by $\mu$ and using the third equation of
\eqref{KKTfmuapprox}, we obtain that
\begin{align}\label{bound}
  \sum_{i=1}^n\frac{w^\star_i}{w_i(\mu,v)}+\sum_{i=1}^n\frac{u^\star_i}{u_i(\mu,v)}& \ \le \
  2C+n.
\end{align}
Since $(w^\star,u^\star)\ge 0$ and $(w(\mu,v), u(\mu,v)) >0$, it follows from \eqref{bound} that for all $i$,
\begin{equation}  \label{low-bdd}
 w_i(\mu,v) \ \ge  \ \frac{w^\star_i}{2C+n}, \qquad u_i(\mu,v) \ \ge  \ \frac{u^\star_i}{2C+n}.
\end{equation}
It immediately implies that the $i$th entry of any accumulation point
$w^\diamond$ of $w(\mu,v)$ as $(\mu,v) \xrightarrow[\Xi_C]{} (0,0)$
must be positive whenever $w^\star_i>0$. Since $w^\diamond$
is an optimal solution of \eqref{design} by part (a), we conclude that part (b) holds.
\end{proof}

\end{document}